\documentclass[longauth]{aa}  
\usepackage{graphicx}
\usepackage{txfonts}
\usepackage{color}
\usepackage{xcolor}
\usepackage{soul}
\sethlcolor{orange}
\definecolor{byzantine}{rgb}{0.74, 0.2, 0.64}

\usepackage{hyperref}
\usepackage{placeins} 
\hypersetup{
    colorlinks=true,
    citecolor=blue,
    linkcolor=blue,
    urlcolor=blue,
}
\usepackage{comment}

\defcitealias{Nascimbeni2022}{N22}
\defcitealias{Montalto2021}{M21}

\begin{document} 

   \title{The PLATO field selection process}

   \subtitle{III. Selection of the Prime Sample for the LOPS2 field}

   \author{
   V.~Nascimbeni\thanks{\email{valerio.nascimbeni@inaf.it}} \inst{\ref{inst1},\ref{inst20}}$^{\href{https://orcid.org/0000-0001-9770-1214}{\includegraphics[scale=0.5]{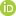}}}$ \and
   G.~Piotto\inst{\ref{inst20},\ref{inst2},\ref{inst1}}$^{\href{https://orcid.org/0000-0002-9937-6387}{\includegraphics[scale=0.5]{orcid.jpg}}}$ \and
    V.~Granata\inst{\ref{inst20},\ref{inst1}\href{https://orcid.org/0000-0002-1425-4541}{\includegraphics[scale=0.5]{orcid.jpg}}} \and
   S.~Marinoni\inst{\ref{inst10},\ref{inst11}}  \and
   P.~M.~Marrese \inst{\ref{inst10},\ref{inst11}} \and
   M.~Montalto  \inst{\ref{inst13}}\and
   J.~Cabrera \inst{\ref{inst5}} \and
   C.~Aerts \inst{\ref{inst3}} \and
   G.~Altavilla\inst{\ref{inst10},\ref{inst11}} \and
   K.~Belkacem \inst{\ref{inst21}} \and
   S.~Benatti \inst{\ref{inst19}}\and
   M.~Bergemann \inst{\ref{inst26}} \and
   A.~B\"orner \inst{\ref{inst4}} \and
   G. Covone \inst{\ref{inst23},\ref{inst24},\ref{inst25}} \and
   M.~Deleuil \inst{\ref{inst6}}\and
   S.~Desidera \inst{\ref{inst1}} \and 
   L.~Gizon  \inst{\ref{inst7}} \and
   M.~J.~Goupil \inst{\ref{inst8}} \and
   M.~G\"unther \inst{\ref{inst9}} \and
   A.~M.~Heras \inst{\ref{inst9}}\and
   L.~Malavolta \inst{\ref{inst2},\ref{inst1}} \and
   J.~M.~Mas-Hesse \inst{\ref{inst12}}\and
   D.~Nardiello \inst{\ref{inst2},\ref{inst1}} \and
   H.~P.~Osborn \inst{\ref{inst14}} \and
   I.~Pagano \inst{\ref{inst13}}\and
   C.~Paproth \inst{\ref{inst4}}\and
   D.~Pollacco \inst{\ref{inst15}}\and
   L.~Prisinzano \inst{\ref{inst19}}\and
   R.~Ragazzoni \inst{\ref{inst2},\ref{inst1}}  \and
   G.~Ramsay \inst{\ref{inst16}}\and
   H.~Rauer \inst{\ref{inst22},\ref{inst18}}\and
   S.~Udry \inst{\ref{inst17}} \and
   T.~Zingales \inst{\ref{inst2},\ref{inst1}} 
          }

   \institute{
   INAF -- Osservatorio Astronomico di Padova, vicolo dell'Osservatorio 5, 35122 Padova, Italy \label{inst1} \and
      Centro di Ateneo di Studi e Attivit\`a Spaziali ``Giuseppe Colombo'' (CISAS), Universit\`a degli Studi di Padova, via Venezia 1,  35131 Padova \label{inst20} \and
   Dipartimento di Fisica e Astronomia ``Galileo Galilei'', Universit\`a degli Studi di Padova, Vicolo dell'Osservatorio 3, 35122 Padova, Italy \label{inst2} \and
   INAF -- Osservatorio Astronomico di Roma, Via Frascati, 33, 00078 Monte Porzio Catone (RM), Italy \label{inst10} \and
   SSDC-ASI, Via del Politecnico, snc, 00133 Roma, Italy \label{inst11} \and
   INAF -- Osservatorio Astrofisico di Catania, Via S. Sofia 78, 95123, Catania, Italy \label{inst13} \and
   Deutsches Zentrum f{\"u}r Luft- und Raumfahrt (DLR), Institut f{\"u}r Weltraumforschung, Rutherfordstra{\ss}e 2, 12489 Berlin-Adlershof, Germany \label{inst5} \and   
   Institute of Astronomy, KU Leuven, Celestijnenlaan 200D, 3001, Leuven, Belgium \label{inst3}\and
      LIRA, Observatoire de Paris, Universit\'e PSL, CNRS, Sorbonne Universit\'e, Universit\'e Paris Cit\'e, CY Cergy Paris Universit\'e, 92190 Meudon, France \label{inst21} \and
   INAF–Osservatorio Astronomico di Palermo, Piazza del Parlamento, 1, I-90129, Palermo, Italy \label{inst19} \and
   Max Planck Institute for Astronomy, Koenigstuhl 17, 69117, Heidelberg \label{inst26} \and
   Deutsches  Zentrum  f\"ur  Luft-  und  Raumfahrt  (DLR),  Institut  f\"ur  Optische  Sensorsysteme, Rutherfordstra{\ss}e  2, 12489 Berlin-Adlershof, Germany \label{inst4} \and   
   Department of Physics ``Ettore Pancini'', University of Naples Federico II, Naples, Italy\label{inst23} \and
   INFN section of Naples, Via Cinthia 6, 80126 Napoli, Italy\label{inst24} \and
   INAF, Osservatorio Astronomico di Capodimonte, Salita Moiariello 16, I-80131 Naples, Italy\label{inst25} \and
   Aix-Marseille Universit\'e, CNRS, CNES, Laboratoire d’Astrophysique de Marseille, Technop\^{o}le de Marseille-Etoile, 38, rue Fr\'ed\'eric Joliot-Curie, 13388 Marseille cedex 13, France \label{inst6} \and
   Max-Planck-Institut f\"ur Sonnensystemforschung, Justus-von-Liebig-Weg~3, 37077~G\"ottingen, Germany \label{inst7}\and
      LESIA, CNRS UMR 8109, Universit\'e Pierre et Marie Curie, Universit\'e Denis Diderot, Observatoire de Paris, 92195 Meudon, France\label{inst8}\and
   European Space Agency (ESA), European Space Research and Technology Centre (ESTEC), Keplerlaan 1, 2201 AZ Noordwijk, The Netherlands \label{inst9} \and   
   Centro de Astrobiolog\'{\i}a (CSIC--INTA), Depto. de Astrof\'{\i}sica, 28692 Villanueva de la Ca\~nada, Madrid, Spain \label{inst12}\and
   Center for Space and Habitability, University of Bern, Bern, Switzerland. \label{inst14} \and
   Department of Physics, University of Warwick, Gibbet Hill Road, Coventry CV4 7AL, UK \label{inst15} \and
   Armagh Observatory \& Planetarium, College Hill, Armagh, BT61 9DG, UK \label{inst16} \and
   Deutsches Zentrum f{\"u}r Luft- und Raumfahrt, Markgrafenstra{\ss}e, 37, 10117 Berlin, Germany  \label{inst22} \and
   Institute of Geological Sciences, Freie Universit\"at Berlin, Malteserstra{\ss}e 74-100, 12249 Berlin, Germany \label{inst18} \and
   Observatoire de Gen\`eve, Universit\'e de Gen\`eve, Chemin Pegasi 51, 1290 Sauverny, Switzerland \label{inst17} 
   }

   \date{Received 5 March 2026 / Accepted 23 April 2026}

  \abstract{The PLanetary Transits and Oscillations of stars (PLATO) mission will begin its four-year nominal mission in early 2027 by monitoring its Long-duration Observation Phase field at South (LOPS2) for at least two years continuously. The primary aim of PLATO is a very ambitious and challenging one: the discovery of Earth-like planets in the habitable zone of nearby and bright solar analogues. To this purpose, the PLATO Mission Consortium, through its Ground-based Observing Program, will perform the follow-up needed to confirm part of the candidate planets photometrically detected by PLATO and measure their masses through radial velocity curves. For the LOPS2, the Ground-based Observing Program is committed (as part of the PLATO mission) to follow-up the candidate exoplanets discovered orbiting the $15\,000$ high-quality target subset of the PLATO Input Catalog (PIC) known as the Prime Sample. The Prime Sample will be made public nine months before launch in the context of the first Guest Observer call for proposals to be issued by the European Space Agency. Here, we present the quantitative metrics and thresholds defined to select and prioritize the Prime Sample. Our method is perfectly general and suitable to rank any list of stars surveyed for transiting planets. We also describe the astrophysical properties of the LOPS2 Prime Sample, both in a statistical sense and for some specific targets of interest.}


   \maketitle
\nolinenumbers

\section{Introduction}\label{sec:introduction}

More than thirty years have passed since the discovery of the first exoplanets \citep{Wolszczan1992,Mayor1995} and about twenty since the start of space-based, wide-field photometric surveys specifically designed to detect transiting exoplanets, namely CoRoT \citep{Auvergne2009}, \emph{Kepler} \citep{Borucki2010}, its successor \emph{K2} \citep{Howell2014}, and the still operating TESS \citep{Ricker2015}. Despite the discovery of thousands of planets and the enormous scientific exploration of related science cases, no Earth-sized planet orbiting in the habitable zone (HZ) of a solar analogue has yet been discovered. Throughout this paper, we define as ``solar analogues'' main-sequence G-type stars (following the approximate $T_\mathrm{eff}$ definition given by \citealt{deStrobel1996} and \citealt{Soderblom1998}) and as ``Earth-like'' planets Earth-sized planets in the HZ of solar-type (FGK) stars.

The discovery of such ``Goldilocks'' planets is the most important scientific goal of the ESA mission PLATO (PLanetary Transits and Oscillation of stars; \citealt{Rauer2025}), a satellite hosting an array of 26 wide-field telescopes to be launched in January 2027 to the L2 Lagrangian point. PLATO will exploit the combination of a very large field of view ($\sim 2149$~deg$^2$), an exquisite photometric precision and accuracy (down to a 10~ppm level noise floor), and the continuous monitoring of the so-called LOP (Long-duration Observing Phase; \citealt{Nascimbeni2022}) fields for at least two years each. The mission aim is to increase our discovery space for rocky planets up to the HZ of main-sequence solar-type dwarfs, including solar twins. The confirmation of these candidate planets will be possible thanks to the brightness of their host stars, which enables ground-based follow-up observations for planet candidate confirmation, including ultra-high-precision radial velocities (RV). The latter will also be crucial in measuring the mass of these planets, allowing us to constrain their physical composition and inner structure. The expected yield of newly discovered PLATO planets around bright stars ($V<11$) is about 1200 \citep{Rauer2025}, of which 10-50 will be Earth-like planets (depending on different assumptions on the actual noise properties and on their occurrence rate $\eta_\oplus$; \citealt{Heller2022,Matuszewski2023,Rauer2025}). HZ rocky planets will require a significant effort not only in terms of detection, verification and validation, but also and above all in the subsequent confirmation phase.
In addition, for the brightest targets ($V\lesssim11$), the light curves will be used for asteroseismology studies, with the possibilities to measure stellar masses, radii, and ages with uncertainties of 15\%, 2\%, and 10\% respectively \citep{Goupil2024}.
To achieve this, the primary challenge lies in overcoming the limitations of current models (e.g., \citealt{Tayar2022}). However, the inclusion of seismic constraints is a key factor that substantially enhances the reliability of the results (e.~g., \citealt{Lebreton2014}). Additionally, the goal of achieving a 10\% uncertainty in stellar age determination —applicable to solar-type stars without a convective core— is expected to be met by the conclusion of a two-year observational campaign, coinciding with the finalization of the catalog. This ambition assumes that systematic biases and errors will have been significantly reduced through advancements in both methodology and theory (e.g., consensus-based reference solar chemical composition, constraints on helium abundance, etc.). This underscores the critical importance of assembling a representative sample of solar-type stars spanning a broad range of parameters, including effective temperature, luminosity (or radius), rotation, and surface chemical composition. Such a sample will enable the testing and validation of new theoretical and empirical formulations for convection and helium abundance. Ultimately, the creation of a PLATO Legacy sample —comprising stars that are exceptionally well-characterized both seismically and non-seismically— is essential, particularly from the perspective of stellar physics.

The ground-based follow-up will be part of the PLATO core mission, and will be coordinated by the PLATO Consortium through its Ground-based Observation Program (GOP; \citealt{Mowlavi2024}; a paper describing in detail the GOP workflow is currently in prep.). The GOP will not monitor the full yield of candidate planets found in the PLATO Input Catalog (PIC) main scientific targets  (i.~e., the $217\,741$ stars contained in the tPIC\footnote{Throughout this paper, we will refer to ``tPIC'' as the current version of the catalog, labeled \texttt{LOPS2PIC2.2-t}.} sub-catalog, where ``t'' stands for ``targets''; \citealt{Montalto2021,Montalto2026}). It will focus instead on a bright and carefully chosen subset of the tPIC, called Prime Sample (PS).  The scope of this paper is to describe the metrics and the selection thresholds adopted to extract the PS for the first LOP field that will be observed by PLATO for at least two years, i.e., the so-called ``LOPS2'' field \citep{Nascimbeni2025}. The duration of this first long pointing could be extended up to four years of the nominal mission, depending on the number of Earth-like planet candidates detected during the first two years. 

This paper is organized as follows. We define the requirements for the PLATO PS in Section~\ref{sec:primesample}. We develop two different metrics to select the PS in Section~\ref{sec:metrics}, and discuss how to apply them and how to define the corresponding selection thresholds for the current version of the PLATO Input Catalog in Section~\ref{sec:selection}. Finally, we describe the astrophysical properties of the PS in Section~\ref{sec:properties}, summarize the ongoing preparatory observations with 4MOST in Section~\ref{sec:4most}, and highlight some prospects for the future and opportunities for the general community in Section~\ref{conclusions}. A glossary of the most common acronyms used throughout this work is compiled in Table~\ref{table:glossary}.

\begin{table}\centering\small
\caption{Glossary of acronyms used throughout this article.}
\begin{tabular}{lp{6cm}}
\hline\hline
Acronym & Description \\
\hline  \noalign{\smallskip}
FCAM & PLATO Fast camera \\
FOV & Field Of View \\
GO & PLATO Guest Observer program\\
GOP & PLATO Ground-based Observation Program\\
HZ & Habitable Zone \\
LOP & Long-duration Observation Phase \\
LOPN & LOP field North\\
LOPN1 & Current LOPN proposal \citep{Nascimbeni2022}\\
LOPS & LOP field South\\
LOPS2 & First LOPS field \citep{Nascimbeni2025}\\
NCAM & PLATO Normal camera \\
NSR & Noise-to-Signal ratio, $(\textrm{S/N})^{-1}$ \\
PIC & PLATO Input Catalog \citep{Montalto2026} \\
PLATO & PLAnetary Transits and Oscillations of stars \citep{Rauer2025} \\
PMC & PLATO Mission Consortium \\
PPT & PLATO Performance Team \\
PropS & PLATO Proprietary Sample \\
PS & Prime Sample \\
PSWT & ESA PLATO Science Working Team \\
RV & Radial velocity \\
SMP & ESA PLATO Science Management Plan \\
S/N & Signal-to-Noise ratio \\
SOP & Step and stare Observation Phase \\
SciRD & PLATO Scientific Requirements Document \\
SRJD & PLATO Scientific Requirements Justification Document \\
tPIC & target PIC subset, \texttt{LOPS2PIC2.2\_t} \citep{Montalto2026} \\
TPT & Target Programming Tool \\
\hline
\end{tabular}\label{table:glossary}
\end{table}

\section{The PLATO Prime Sample}\label{sec:primesample}

The PS is formally defined in the ESA PLATO Science Management Plan\footnote{\url{https://www.cosmos.esa.int/documents/1635684/1636120/PLATO_Science_Management_Plan_ESA.pdf}} (SMP) as follows:

\begin{quote}\emph{
A ``prime sample'' will be defined by the Plato Mission Consortium, consisting of PIC targets to be observed with high PLATO accuracy. The ``prime sample'' with the targets in the first LOP sky field will be defined nine months before launch and updated six months before every satellite sky field pointing. Ground-based observations will be performed for candidates within this sample during the course of the mission.}
\end{quote}

In other words, the PS is strongly follow-up oriented and is supposed to include the best available targets in terms of both detection and confirmation efficiency. 
In the SMP, it is also specified that ``the size of the Prime Sample will not be larger than 20,000 stars''. 
The PLATO Science Working Team (PSWT) decided to reserve $15\,000$ PS targets for the first Long-duration Observation Phase (LOP) field (LOPS2), reserving the right to allocate the remaining $5\,000$ additional PS targets for other possible LOP fields, even though the choice of the next LOP field(s) to be pointed to will be made after launch. As it will be explained in Section~\ref{sec:selection}, the actual number of PIC stars flagged as PS targets will be slightly larger than $15\,000$, because the PIC includes a safety margin around LOPS2 to take into account pointing errors and camera misalignment.

Two facts about the PS are of particular interest for the astrophysical community at large. Firstly, the PS stars cannot be targeted by any GO proposal, regardless of the proposed science case. For this reason, the PS list will be released to the community at the time of the first public release of the PIC (PIC2.2) and available for the preparation of the GO proposals. Secondly, while in general the L0 and L1 data products (raw and calibrated imagettes, light curves and centroid curve, respectively, as defined in Section~8.2 of \citealt{Rauer2025}) for most tPIC targets will be released almost immediately after technical validation \citep{Rauer2025}, this is not true for the $15\,000$ PS stars. PS data will be released after the scientific validation of the candidate planets, in any case, not later than one year after the validation of L0/L1.  
An important aspect of the PS is the commitment by the PMC to deliver the Level-3 products for these targets, that is, as the SMP states, the resulting 
``list of confirmed planetary systems, which will be fully characterised by combining information from the planetary transits, the seismology of the planet-hosting stars, and the results of ground-based observations.''

\begin{figure}\centering
    \includegraphics[width=0.9\columnwidth]{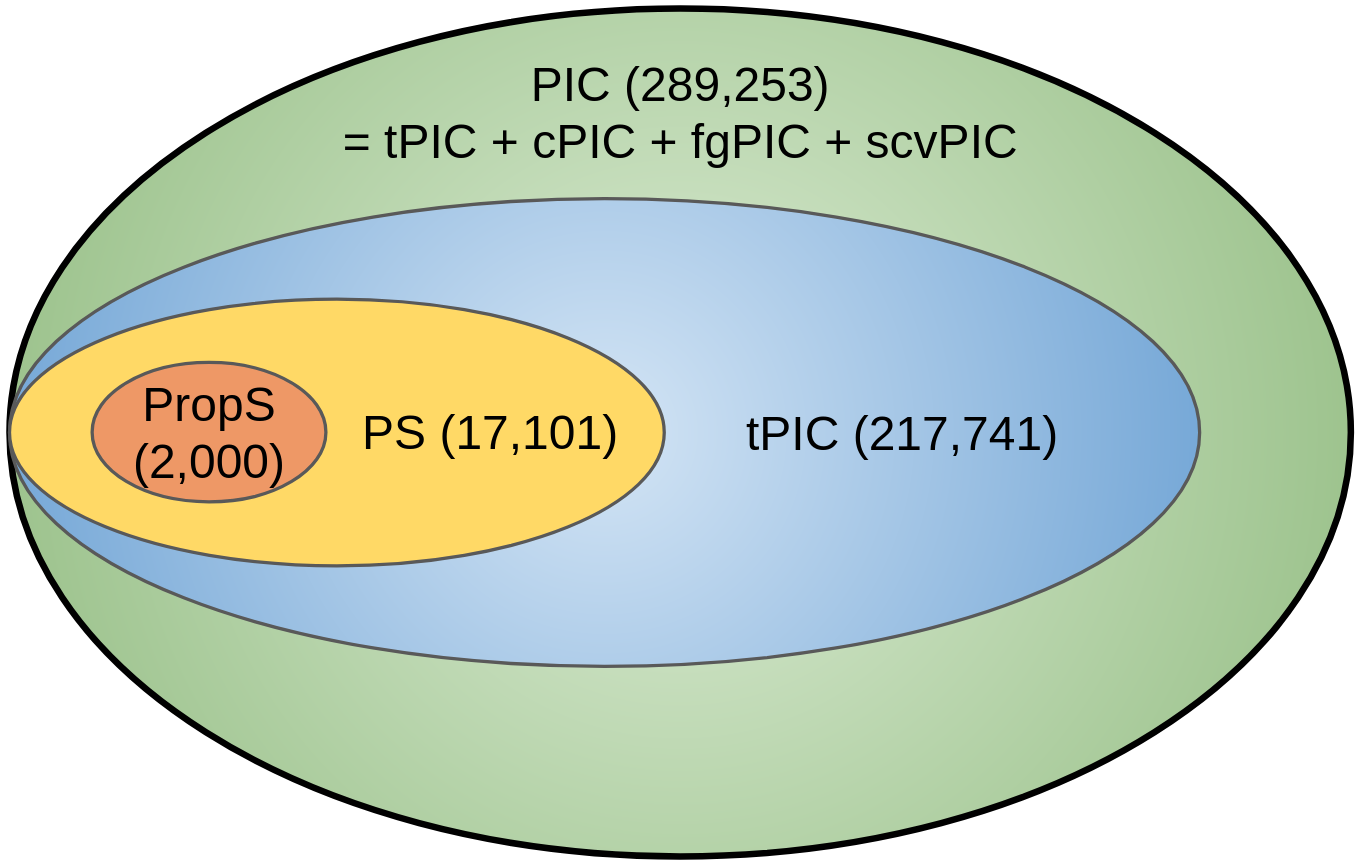}
    \caption{Venn diagram illustrating the main subsets of the Plato Input Catalog (PIC; version 2.2) relevant for this paper: The target PIC (tPIC), the Prime Sample (PS), the Proprietary Sample (PropS). Details in Section~\ref{sec:primesample}.}
    \label{fig:venn}
\end{figure}

For a subset of the PS consisting of a maximum of 2000 stars, the so-called the Proprietary Sample (PropS), data will be released six months after the completion of the follow-up, and, in any case, at the end of the mission in its final archive. The PropS will be the subject of a future article in this series, following the selection of PropS targets, expected after the first quarter
of PLATO observations. A Venn diagram illustrating how the PropS, PS, tPIC, and PIC sets are nested within each other and how many targets they include is shown in Fig.~\ref{fig:venn}.

\begin{table}\centering\small
\caption{Definitions of the main PLATO target stellar samples.}
\begin{tabular}{ccp{1.55cm}cc}
\hline\hline
sample & SpT & lum.~class & mag. limit & noise \\ \hline \noalign{\smallskip}
P1 & F5-K7 & dwarfs and subgiants & $V<11$ & $<50$~ppm \\ 
P2 & F5-K7 & dwarfs and subgiants & $V<8.5$ & $<50$~ppm \\ 
P4 & M & dwarfs & $V<16$ & \dots \\ 
P5 & F5-K7 & dwarfs and subgiants & $V<13$ & \dots \\ \hline
\end{tabular}\label{tab:samples}
\tablefoot{The rows give: the name of the PLATO target stellar sample, the spectral type, the luminosity class, the limiting magnitude in the $V$ band, and the noise limit in ppm in one hour.}
\end{table}

\section{The base metrics}\label{sec:metrics}

Unlike the approach previously adopted to select the main tPIC subsamples named P1-P2-P4-P5 defined by \citet{Nascimbeni2022} and \citet{Montalto2026} and summarized in Table~\ref{tab:samples}, we decided to base the whole PS selection process on what we call \emph{metrics}, i.~e., on one-dimensional merit functions of the stellar parameters designed to capture the scientific value of a given target with a single number. Such metrics have important advantages over simpler approaches based on fixed thresholds:
\begin{itemize}
    \item They can be tuned and normalized to match the expected performances of the instrument, for example, by assigning a value of 1 to stars right on the detection threshold for Earth analogs;
    \item They avoid sharp edges in the parameter space as much as possible, in order to be more efficient in optimally selecting only the most suitable targets. Of course, at some point sharp limits due to operational constraints (such as on stellar magnitude) have to be applied to our PS, but their statistical impact is easy to be managed (see also the following point);
    \item They can be easily reproduced in any synthetic sample, to quantify the selection effects at play and therefore to reconstruct the underlying statistical occurrence of planets (a crucial ingredient to properly estimate $\eta_\oplus$);
    \item They not only select but also rank the targets according to their value as a sorted list in a natural way. Consequently, replacing a target excluded for any reason or changing the sample size at any time is a trivial process.
\end{itemize}

The latter advantage is obvious not only for the PS, but also for the whole sample of tPIC targets at large, since the technical requirements of PLATO will sometimes impose an automated choice between targets that are mutually exclusive. Indeed, as we mention in Section~\ref{conclusions}, one of the metrics presented in the next Section is also currently adopted as a basis for the tPIC target priority ranking listed in the  \texttt{tPICScientificRanking} column in PIC2.2.

For a metric to be effective, it must be possible to define it as a simple analytical function of parameters that are available for all the stars in our sample, and derived as homogeneously as possible. In our case, this forces us to rely on quantities tabulated in the tPIC, such as the stellar radius $R_\star$, mass $M_\star$, effective temperature $T_\mathrm{eff}$ and the noise-to-signal ratio (NSR) estimated by the Plato Performance Team (PPT) through the PINE code, \citealt{Borner2024}; the NSR is defined as the inverse of the S/N).

We can identify two main stages in the process leading to the confirmation of a newly discovered planet: the photometric detection of a candidate transit from the PLATO light curves, and the subsequent ground-based follow-up necessary to exclude any false-positive scenario and measure the planet parameters. In the following Sections (\ref{sec:m1m2} and \ref{sec:r1r2}, respectively), we develop two families of signal-to-noise ratio (S/N) based metrics to represent each stage.

\subsection{The photometric detection metric $\mathcal{M}$}
\label{sec:m1m2}

\begin{figure*}
\centering
    \includegraphics[width=0.9\columnwidth]{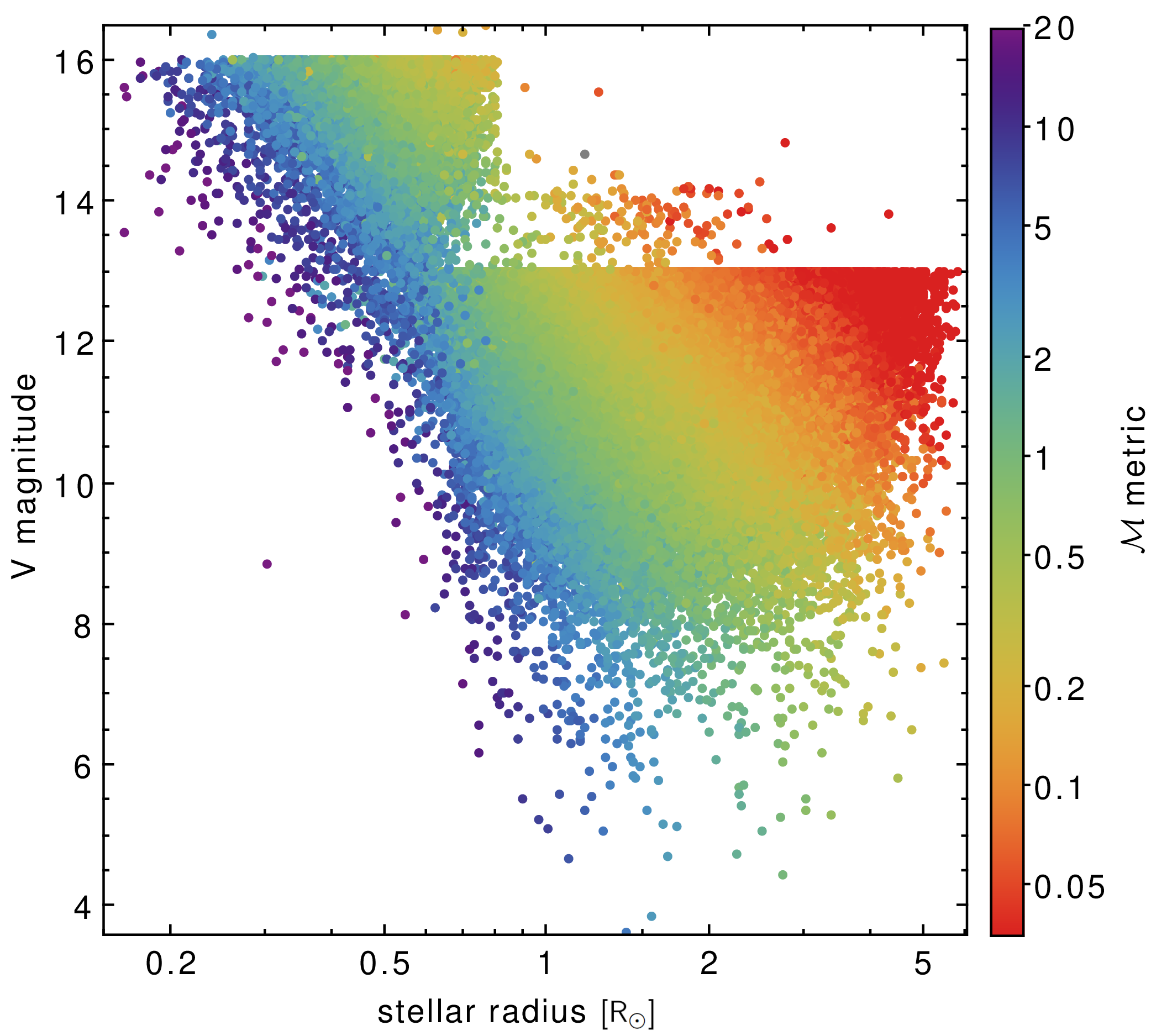}  \hspace{5mm}  \includegraphics[width=0.9\columnwidth]{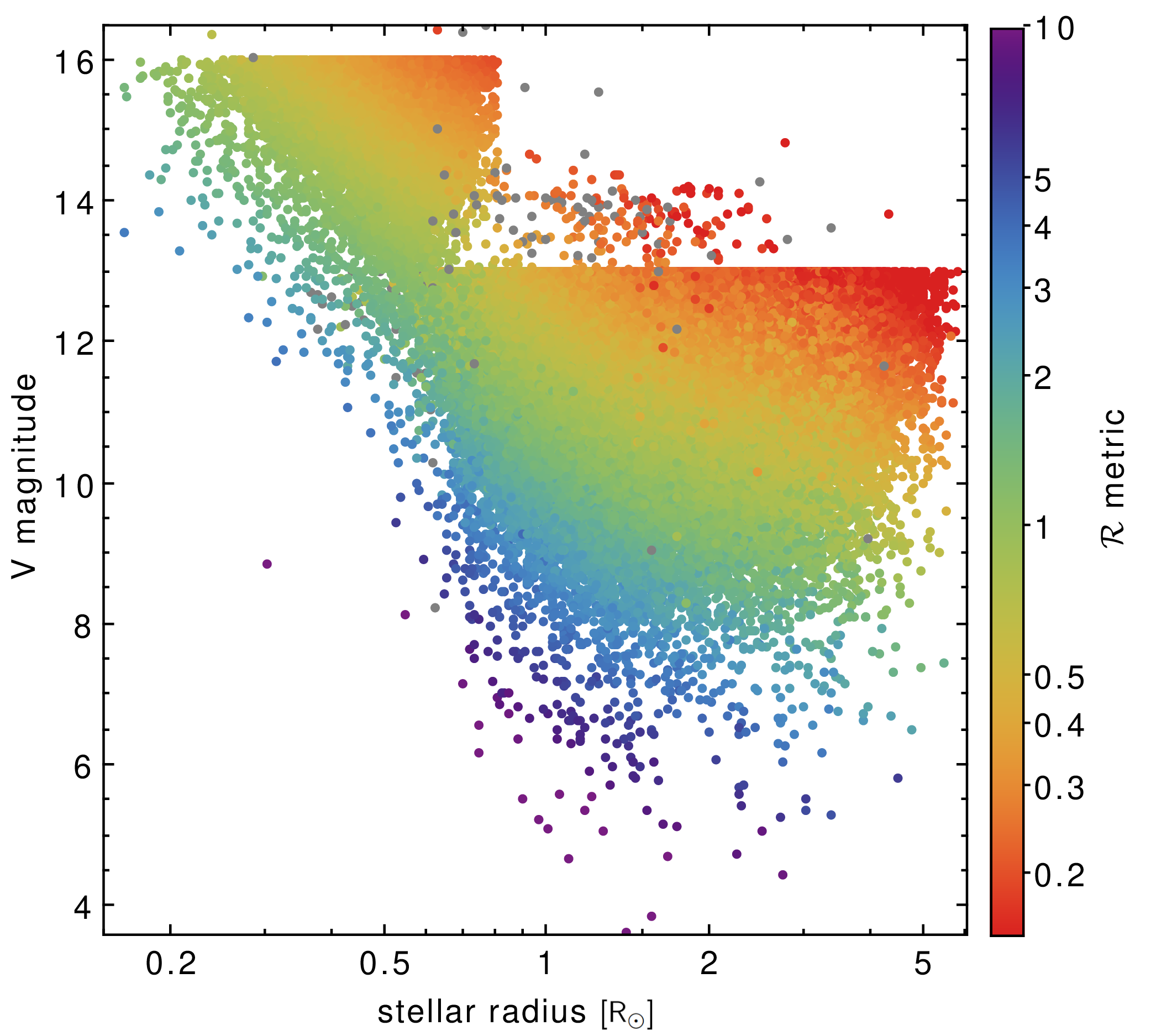}
\caption{Metrics $\mathcal{M}$ and $\mathcal{R}$ (as defined in Sections~\ref{sec:m1m2} and \ref{sec:r1r2}) applied to our stellar sample. \emph{Left plot:} the whole sample of $217\, 741$ stars in tPIC2.2 plotted as a function of stellar radius $R_\star$ and $V$ magnitude and color-coded according to the value of $\mathcal{M}$ metric (defined in Section~\ref{sec:m1m2}). The sharp cuts at $V=13$ and $V=16$ are due to the magnitude requirements of samples P5 and P4, respectively (Table~\ref{tab:samples}). \emph{Right plot:} same, but for $\mathcal{R}$  (defined in Section~\ref{sec:r1r2}). Note that the color scale does not span the same numerical range.}
\label{fig:metrics}
\end{figure*}

The first metric we introduce ($\mathcal{M}$) is based on the expected S/N of the photometric detection of a HZ Earth-sized planet by PLATO, but it can be trivially generalized to any other instrument by properly adjusting the noise computation. Let us split and calculate separately the signal and the noise part:
\begin{equation}
    \textrm{S/N} = \frac{\mathcal{S}_\mathrm{phot}}{\mathcal{N}_\mathrm{phot}}\textrm{ .}
\end{equation}
The signal  $\mathcal{S}_\mathrm{phot}$ has to be expressed per time unit, since each PLATO field will be observed for a finite time range ($\ge 2$~years for the LOP phase). Assuming purely Poissonian white noise and approximating the transit shape as a rectangular box of width $T_\mathrm{tot}$, the transit depth $\delta$ has to be scaled by the square root of both the transit duration $T_\mathrm{tot}$ and the number of transits gathered $N_\mathrm{tran}$ over the observation span:
\begin{equation}\label{eq:sphot1}
    \mathcal{S}_\mathrm{phot} = \delta \times \sqrt{T_\mathrm{tot}} \times \sqrt{N_\mathrm{tran}}\textrm{ .}
\end{equation}
Neglecting the limb darkening effect, we can approximate the transit depth as $\delta\approx(R_\mathrm{p}/R_\star)^2$, where $R_\mathrm{p}$ and $R_\star$ are the planetary and stellar radius, respectively. The number of transits $N_\mathrm{tran}$ is, of course, proportional to the inverse of the orbital period $P$. Then, following \citet{Winn2010}, for circular orbits, we get an expression for $T_\mathrm{tot}$ as a function of $P$, radius ratio $k=R_\mathrm{p}/R_\star$, impact parameter $b$, orbital inclination $i$, semi-major axis $a$ that can be further simplified by assuming small planets ($k\ll1$), central transits ($b\simeq 0$) and neglecting very close-in planets ($a/R_\star \gg 1$):
\begin{equation}
    T_{\rm tot} =
\frac{P}{\pi}\arcsin\left[\frac{R_\star}{a}
  \frac{\sqrt{(1+k)^2-b^2}}{\sin i} \right ] \approx \frac{P}{\pi}\frac{R_\star}{a}\textrm{ .}
\end{equation} 
If we substitute this expression in Eq.~(\ref{eq:sphot1}), we get
\begin{equation}
    \mathcal{S}_\mathrm{phot} = \left ( \frac{R_\mathrm{p}}{R_\star} \right )^2 \sqrt{\frac{PR_\star}{\pi a}} \sqrt{\frac{1}{P}}  \textrm{ ,}
\end{equation}
and, by simplifying and collecting all the constants, we get a function of $R_\star$ and $a$ only:
\begin{equation}\label{eq:sphot2}
    \mathcal{S}_\mathrm{phot} \propto R_\star^{-3/2}a^{-1/2} \textrm{ .}
\end{equation}
If we assume a constant period for the planet (and therefore a constant $a$), we could derive a metric which is essentially identical to that adopted to prioritize the TESS targets \citep{Stassun2018} or, with only minor modifications, to select the PLATO LOP fields \citep{Nascimbeni2022}. However, in this context, we focus specifically on the detection of HZ planets. So, instead of fixing $a$ in Eq.~(\ref{eq:sphot2}), we assume a semi-major axis corresponding to the irradiation level of the Earth-Sun system as defined by \citet{Kasting1993}:
\begin{equation}\label{eq:ahz}
    a_\mathrm{HZ} \textrm{ [au]} = 1.0\times\sqrt{L_\star/L_\odot} = \left ( \frac{T_\mathrm{eff}}{T_\mathrm{teff,\odot}} \right ) ^2 \frac{R_\star}{R_\odot} \textrm{ ,}
\end{equation}
where we assumed $T_\mathrm{teff,\odot}=5778$~K as the effective temperature $T_\mathrm{eff}$ of the Sun. We are aware that this is a strong assumption, as the position and width of the HZ is expected to depend in a very complex way as a function of both stellar parameters and of the atmospheric composition and structure of the planet \citep{Kasting1993,Kopparapu2013}. Still, the metric we aim to design has to be kept as simple as possible in order to be effective (as explained in Section~\ref{sec:metrics}).

In solar units, the signal from Eq.~\ref{eq:sphot2} can then be expressed as:
\begin{equation}\label{eq:sphot3}
    \mathcal{S}_\mathrm{phot} \propto R_\star^{-2}\,T_\mathrm{eff}^{-1} \textrm{ .}
\end{equation} 

On the other hand, again in the white noise regime, the $\mathcal{N}_\mathrm{phot}$ term can be estimated as the inverse of the square root of the flux density $F_\star$ in the photometric band of interest; in terms of magnitude $m$, this implies 
\begin{equation}\label{eq:nphot}
    \mathcal{N}_\mathrm{phot}\propto\frac{1}{\sqrt{F_\star}} = \frac{1}{\sqrt{10^{-0.4\, m_\star}}} \textrm{ .}
\end{equation}
The metric $\mathcal{M}$ we are searching for can finally be computed by putting together Eq.~(\ref{eq:sphot3}) and (\ref{eq:nphot}):
\begin{equation}\label{eq:m_general_prop}
    \mathcal{M} \propto \frac{\mathcal{S}_\mathrm{phot}}{\mathcal{N}_\mathrm{phot}} = R_\star^{-2}\,T_\mathrm{eff}^{-1}\, 10^{-0.2\, m_\star} \textrm{ ,}
\end{equation}
which is perfectly general. The same metric can be normalized by expressing the last term as $10^{-0.2\,(m_\star-m_0)}$, where $m_0$ is the magnitude at which the instrument of interest is able to yield a detection for an Earth analogue transit on a solar twin:
\begin{equation}\label{eq:m_general}
    \mathcal{M} = R_\star^{-2}\,T_\mathrm{eff}^{-1}\, 10^{-0.2\, (m_\star-m_0)} \textrm{ .}
\end{equation}
By setting $\mathcal{M}=1$ and expressing the formula above as a function of $m_\star$, we get the limiting magnitude $m_\mathrm{lim}$ corresponding to the detection threshold:
\begin{equation}\label{eq:vlim_m}
    m_{\lim,(\mathcal{M}=1)} = -5\log \left (T_\mathrm{eff}\, R_\star^{2}\right )+m_0 \textrm{ .}
\end{equation}

In the specific case of PLATO, $\mathcal{N}_\mathrm{phot}$ is already estimated for each PIC target by taking into account a wide set of random and systematic noise sources \citep{Borner2024} as noise-to-signal ratio (NSR) expressed in ppm in one hour ($\textrm{ppm}/\!\sqrt{[h]}$) and provided in the \texttt{BOLrandomSysNSRNCAM\_T} column. Since, on a LOP field, PLATO will be able to robustly detect an 84-ppm deep Earth-twin transit around a Sun-like star with an NSR equal to or smaller than 80~ppm in one hour \citep{Smith2025,Czismadia2013,Csizmadia2023}, we can assume that value as a reference to normalize our metric:
\begin{equation}\label{eq:m}
   \mathcal{M} = \frac{\textrm{80 ppm}}{\textrm{NSR}}\left ( \frac{R_\star}{R_\odot} \right ) ^ {-2}\left ( \frac{T_\mathrm{eff}}{T_\mathrm{teff,\odot}} \right ) ^{-1} \textrm{.}
\end{equation}
In other words, the $\mathcal{M}>1$ condition selects stars for which PLATO will be able to detect transits of HZ Earth-sized planets. 

The value of $\mathcal{M}$ for every target in the tPIC catalog is plotted with a color scale in the left panel of Fig.~\ref{fig:metrics}. As expected,  the highest values of the metric, and therefore the highest priority, are given to small, bright and cool main-sequence stars, while faint subgiants are at the opposite end. The strong dependence on the stellar radius is essentially dominating the metric. Indeed, a linear fit to the whole tPIC data set of stellar parameters (i.~e., over the typical range of late-type solar-type dwarfs) yields approximately $(R_\star^{-2} T_\mathrm{eff}^{-1})\approx R_\star^{-2.2}$. In other words, leaving the noise level equal, a K2V star gives a $\sim 4.5$ times higher metric (and therefore $4.5\times$ more priority) than an F5V.

\subsection{The follow-up metric: $\mathcal{R}$}\label{sec:r1r2}

The second metric we need to define, called $\mathcal{R}$, is related to the ease of follow-up from the ground, again quantified as an S/N ratio. Since the radial velocity (RV) follow-up will be the most time- and resource-consuming part of the GOP activities, $\mathcal{R}$ must be based on the S/N achievable with the RV technique. In analogy with the approach of the previous Section, let us consider separately the two factors:
\begin{equation}
    \textrm{S/N} = \frac{\mathcal{S}_\mathrm{rv}}{\mathcal{N}_\mathrm{rv}}\textrm{ .}
\end{equation}

The signal to be measured with the RV technique when following up a planet on a circular orbit with period $P$ around a star of mass $M_\star$ is the Keplerian RV semi-amplitude $K$:
\begin{equation}
K=28.4\textrm{ m s}^{-1}\left (\frac{P}{\textrm{1 yr}} \right )^{-1/3} \left ( \frac{M_\mathrm{p}\sin i}{M_\mathrm{jup}}\right )\left ( \frac{M_\star}{M_\odot}\right )^{-2/3} \textrm{ ,}
\end{equation}
hence: 
\begin{equation}\label{eq:srv1}
    \mathrm{S}_\textrm{rv} \propto M_\star^{-2/3} P^{-1/3} \textrm{ ,}
\end{equation} 
where, again, we can replace $P$ with the orbital period $P_\mathrm{HZ}$ corresponding to the reference HZ position according to \citet{Kasting1993}. From Eq.~\ref{eq:ahz}, and knowing from Kepler's law that for $M_\mathrm{p}\ll M_\star$ we have $P \propto a^{3/2}M_\star^{-1/2}$ and therefore 
\begin{equation}\label{eq:phz}
    P_\mathrm{HZ} \textrm{ [yr]} = \left ( \frac{T_\mathrm{eff}}{T_\mathrm{teff,\odot}} \right ) ^3 \left ( \frac{R_\star}{R_\odot} \right ) ^{3/2} \left ( \frac{M_\star}{M_\odot}\right ) ^{-1/2} \textrm{ .}
\end{equation}
The RV signal then becomes, substituting Eq.~(\ref{eq:phz}) into (\ref{eq:srv1}):
\begin{equation}\label{eq:srv2}
       \mathcal{S}_\textrm{rv} \propto M_\star^{-1/2}\, T_\mathrm{eff}^{-1} \,R_\star^{-1/2}\textrm{ .}
\end{equation}
The noise term $\mathcal{N}_\mathrm{rv}$, in analogy with $\mathcal{N}_\mathrm{phot}$, can be modeled as Poissonian noise in the photometric band of interest:
\begin{equation}
    \mathcal{N}_\textrm{RV} \propto 1/\sqrt{F_\star} \propto 10^{0.2\cdot m}\textrm{ .}
\end{equation}
The metric $\mathcal{R}$ can be finally defined as:
\begin{equation}\label{eq:r_general_prop}
        \mathcal{R} \propto M_\star^{-1/2}\, T_\mathrm{eff}^{-1} \,R_\star^{-1/2}\, 10^{-0.2\, m_\star} \textrm{ ,}
\end{equation}
which, just like Eq.~(\ref{eq:m_general_prop}), is perfectly general and can be normalized by defining $m_0$ as the magnitude at which the instrument of interest is able to yield a detection for an Earth analogue transit on a solar twin:
\begin{equation}\label{eq:r_general}
        \mathcal{R} = \mathcal{S}_\mathrm{rv}/\mathcal{N}_\mathrm{rv} \approx M_\star^{-1/2}\, T_\mathrm{eff}^{-1} \,R_\star^{-1/2}\, 10^{-0.2\, (m_\star-m_0)} \textrm{ .}
\end{equation}
Just as done for $\mathcal{M}$, we can set $\mathcal{R}=1$ and express the formula above as a function of $m_\star$, to get the limiting magnitude $m_\mathrm{lim}$ corresponding to the detection threshold:
\begin{equation}\label{eq:rlim_m}
    m_{\lim,(\mathcal{R}=1)} = -5\log \left (M_\star^{1/2} T_\mathrm{eff}\, R_\star^{1/2}\right )+m_0 \textrm{ .}
\end{equation}

In the specific case of the PLATO follow-up, the vast majority of spectrographs available to GOP will be optical ones, for which the $V$-band magnitude is reasonably representative of the S/N achievable for high-precision RV measurements. Besides that, we can consider the scenario of a HZ Earth-mass planet around a Solar analogue at $V=11$ (corresponding to the P1 sample faint limit) as a reference case, and normalize our metric as follows:
\begin{equation}\label{eq:r}
        \mathcal{R} = 158.5 \cdot \left (\frac{M_\star}{M_\odot} \right  ) ^{-1/2} \left ( \frac{T_\mathrm{eff}}{T_\mathrm{teff,\odot}} \right ) ^{-1} \left ( \frac{R_\star}{R_\odot} \right ) ^{-1/2} 10^{-0.2\cdot V} \textrm{ ,}
\end{equation}
After this normalization, and under our previous assumptions, stars with $\mathcal{R}=1$ should, in principle, be accessible with the same S/N by the RV follow-up, as far as HZ Earth-sized planets are concerned. We emphasize that, while $\mathcal{M}$ is an absolute metric (where $\mathcal{M}>1$ implies photometric detectability by PLATO), $\mathcal{R}$ is merely a relative metric, depending on the amount of resources available to the GOP, on the actual performance of the spectrographs and, finally, on the amount of observing time to be allocated to each target. For these reasons, the actual threshold value to be applied to $\mathcal{R}$ is not 1, but will be instead fine-tuned in Section~\ref{sec:selection} in order to set the size of the PS to exactly $15\,000$ targets, as in the PLATO science requirements.

The value of $\mathcal{R}$ for each target in the tPIC catalog is plotted as a color scale in the right panel of Fig.~\ref{fig:metrics}. As for $\mathcal{M}$, the highest values of the metric $\mathcal{R}$ are associated with small, bright and cool main-sequence stars, while faint subgiants are at the opposite end. However, the dependence of $\mathcal{R}$ on the stellar radius is weaker with respect to $\mathcal{M}$; a linear fit to the whole tPIC data set of stellar parameters yields approximately $(M_\star^{-1/2} T_\mathrm{eff}^{-1}R_\star^{-1/2})\approx R_\star^{-1.4}$. By applying the same comparison discussed for $\mathcal{M}$, and the noise level being equal, a K2V star gives a $\sim 2.2$ times higher metric (and therefore $2.2\times$ more priority) than an F5V.

It is worth mentioning that, while the normalization of Eq.~(\ref{eq:r}) has been devised with optical RV spectrographs in mind, nowadays near-infrared (NIR) spectrographs are becoming increasingly effective in delivering high-precision RVs on late-type stars. We take this into account in Section~\ref{sec:selection}, at the PS selection stage.

As a closing note, we emphasize that $\mathcal{R}$, just like $\mathcal{M}$, does not take into account any source of uncertainty other than pure white photon noise, i.~e., we are neglecting astrophysical effects such as stellar variability, stellar rotation, dependence of the RV precision on stellar parameters, etc. In the brightest tail of our input catalog, in particular, we will be dominated almost certainly by a systematic noise floor due to a combination of instrumental and astrophysical effects (this will have no impact on the PS selection, however, as the bright tail of the tPIC, i.~e., the P2 sample will be forced in the PS anyway; see Section~\ref{sec:selection}). All these factors will be taken into account at the vetting stage, on a star-by-star basis. This also implies that we are not explicitly biasing our input sample towards low-activity and therefore older stars; there is also no explicit bias for binary stars or Galactic stellar populations, as a consequence of our general approach of keeping the sample as unbiased as possible in order to carry out appropriate statistical studies on the PLATO planetary yield. Indirect selection effects (such as those driven by the cuts in spectral type or $R_\star$) can still be at play, but the dedicated modeling of these effects is beyond the scope of this paper.

\begin{figure*}\centering
    \includegraphics[width=0.9\columnwidth]{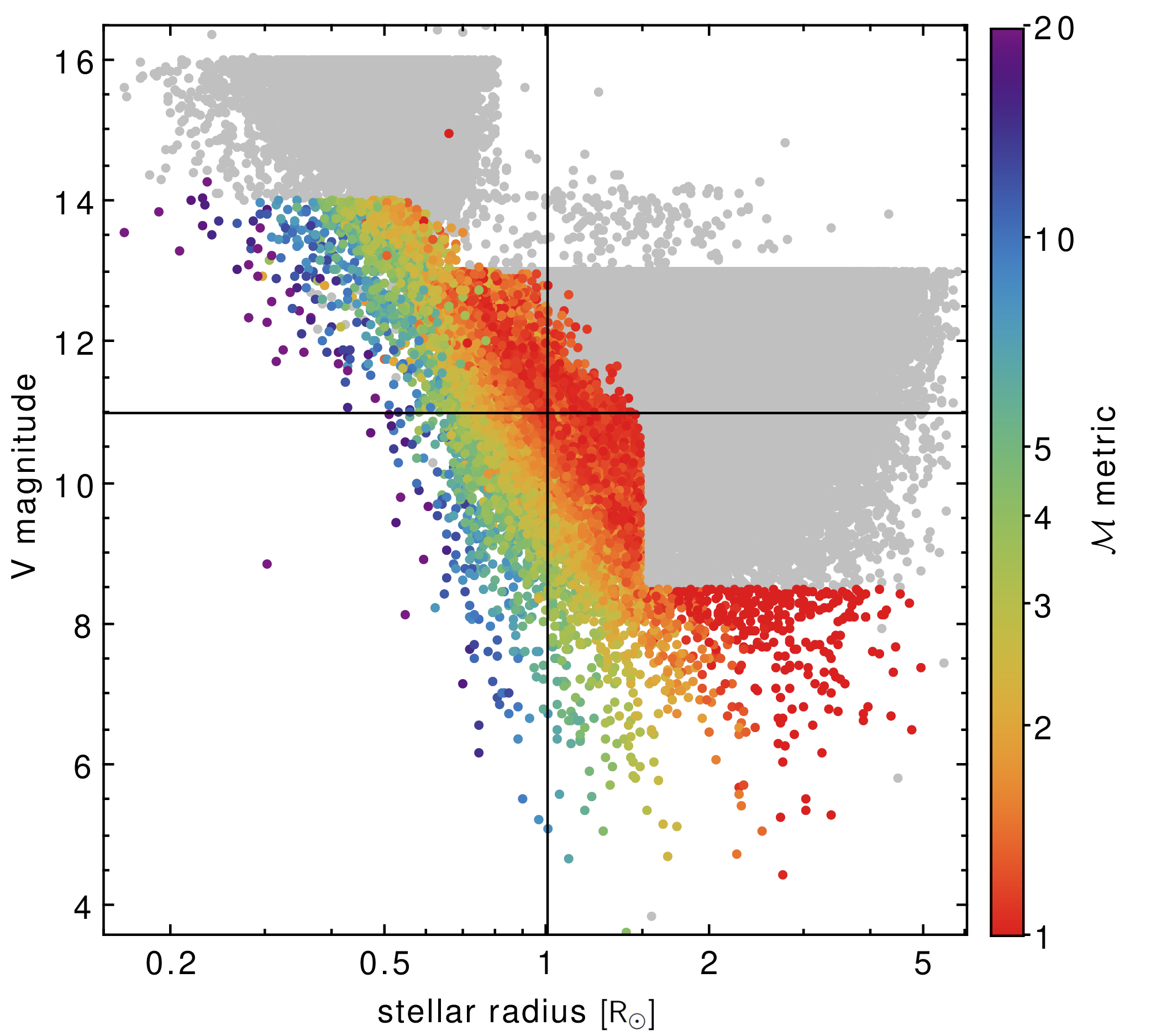}  \hspace{5mm}   \includegraphics[width=0.9\columnwidth]{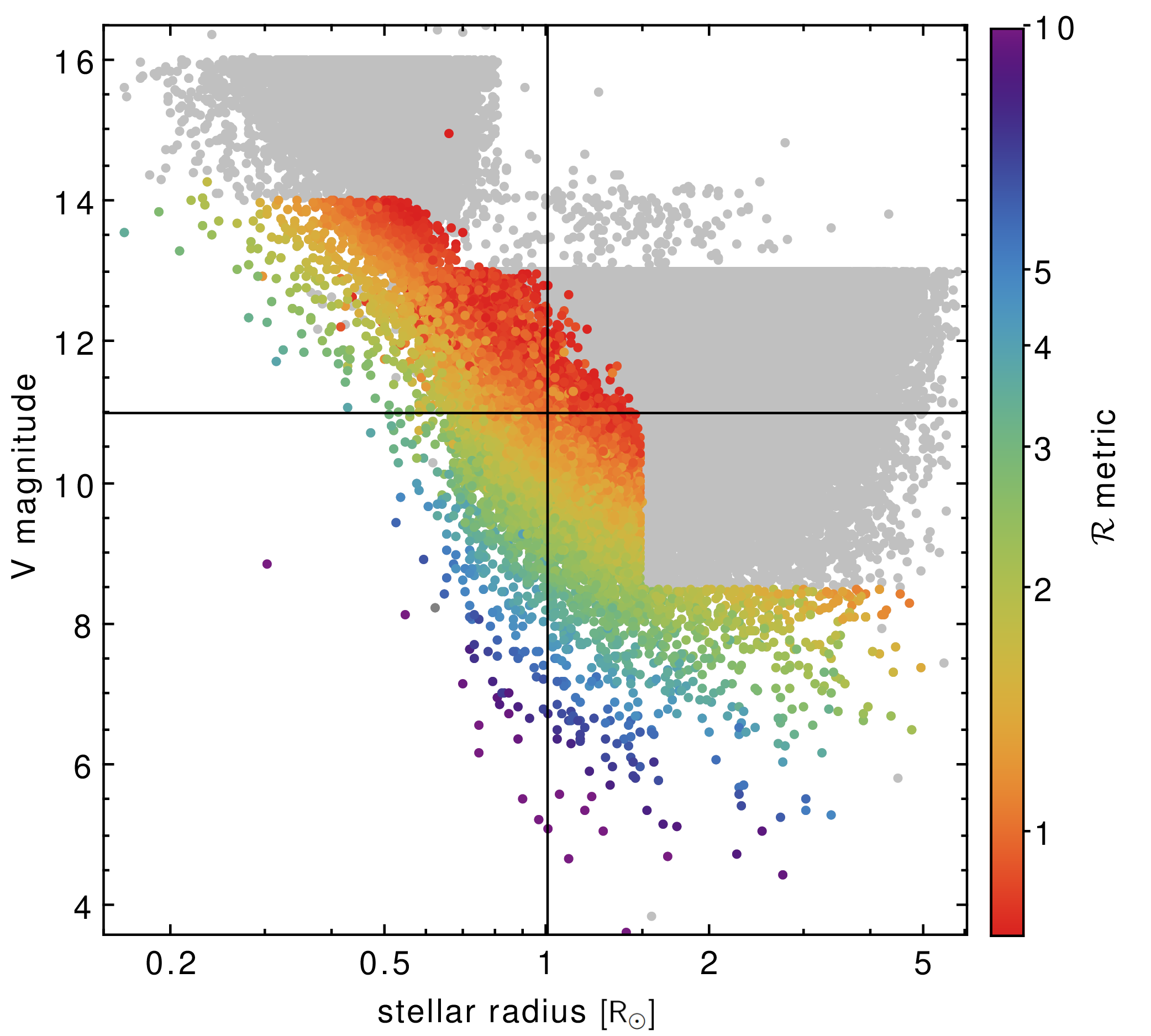}
\caption{Metrics $\mathcal{M}$ (left panel) and $\mathcal{R}$ (right panel), as defined in Sections~\ref{sec:m1m2}, \ref{sec:r1r2} and calculated for the Prime Sample stars (Section~\ref{sec:primesample}). The sharp selection cuts at $V=14$ and $R_\star=1.5$~$R_\odot$ are explicitly set by the inclusion criteria (Eq.~\ref{eq:selcriterion}), while the cut at $V=8.5$ comes from the definition of the P2 subsample (Table~\ref{tab:samples}).}
\label{fig:ps_stats}
\end{figure*}

\section{The selection criteria}\label{sec:selection}

At this point, we are ready to combine the metrics we defined in the previous Section to extract the PS from the tPIC. The PIC2.2  contains a total of $217\,741$ tPIC entries \citep{Montalto2026}, of which only $182\,581$ fall within the LOPS2 nominal footprint. The latter subset can be extracted by selecting stars having $\texttt{BOLnCameraObsNCAM\_T}>0$ and we will refer to it as the ``nominal'' subset from here on. Stars flagged with $\texttt{BOLnCameraObsNCAM\_T}=0$ belong to a safety margin of a few degrees around the nominal field to take into account pointing and repointing errors, rotation offsets and mutual telescope misalignments.

The $\mathcal{M}$ and $\mathcal{R}$ metrics are essential ingredients of our selection recipe for the PS, but we still need a few amendments to force the inclusion or exclusion of specific classes of targets. The PSWT agreed to:
\begin{itemize}
    \item limit the sample at $R_\star<1.5 \, R_\odot$ to exclude large and evolved stars, for which the follow-up could be hindered by fast rotation and/or pulsations;
    \item limit the sample at $V<14$ to exclude faint stars, for which the follow-up could be prohibitively time expensive, regardless of the theoretically predicted NSR.
\end{itemize}
Two classes of targets are then forced into the PS as an exception to the two selection cuts above:
\begin{itemize}
    \item the whole P2 sample (868 stars, 712 of them falling in the nominal field), regardless of the NSR or stellar parameters of the targets. This implies forcing all the tPIC stars with $V<8.5$, recognized as high-priority targets due to their brightness. For instance, all the targets within the LOPS2 footprint and listed in Table D.2 of \citet{Lund2025} as high-priority targets for the detection of solar-like oscillations are included in the PS, and all of them except four would have been rejected by the $R_\star<1.5 \, R_\odot$ requirement. The sole exception is HD~42581 = GJ~229, the only P4 star in the tPIC brighter than $V=8.5$. This target is indeed included in the PS, but due to its high metric values (it does not fall within the nominal LOPS2, however);
    \item all the P4 stars (i.e., M dwarfs; \citealt{Prisinzano2026}) brighter than $H=9$ in the 2MASS system, that is, bright enough to be conveniently observed by NIR spectrographs, recognizing the growing effectiveness of new generation instruments such as NIRPS \citep{Bouchy2025,Vaulato2025}. This metric term selects 275 targets, of which only seven would not have been selected by the previous criteria.
\end{itemize}

Since the selection threshold on $\mathcal{M}$ is set to $\mathcal{M}>1$ by construction (Section~\ref{sec:m1m2}), and since $\mathcal{R}$ is a relative metric as explained in Section~\ref{sec:r1r2}, the threshold on $\mathcal{R}$ is acting here as a free parameter and can be fine-tuned in order to get a total number of $15\,000$ PS stars in the nominal field (as required; see Section~\ref{sec:primesample}). With a compact notation, our selection criteria can be summarized with the following logical expression:
\begin{gather}
    (\mathcal{M}>1 \texttt{ AND } \mathcal{R}>0.73861 \texttt{ AND } R_\star/R_\odot<1.5   \texttt{ AND } V<14) \nonumber \\ 
    \texttt{ OR } (\mathrm{P2}) \texttt{ OR } (\mathrm{P4} \texttt{ AND } H<9)  \textrm { ,} \label{eq:selcriterion}
\end{gather}
where the threshold value of $\mathcal{R}=0.73861$ corresponds to a Solar twin (1~$M_\odot$, 1~$R_\odot$, 5778~K) at magnitude $V\simeq 11.66$. ``P2'' and ``P4'' are boolean flags marking the membership of a target to the P2 or P4 sample, \texttt{AND}/\texttt{OR} are the usual boolean operators. The $\mathcal{M}$ and $\mathcal{R}$ metrics are calculated as in Eq.~(\ref{eq:m}) and (\ref{eq:r}), respectively.
For those 547 stars for which one or more ingredients of the metric are missing ($R_\star$, $M_\star$ and/or $T_\mathrm{eff}$) the value of $\mathcal{M}$ and $\mathcal{R}$ is assigned as the average value of $\mathcal{M}$ and/or $\mathcal{R}$ of their subset of belonging (P1, P2, P4 or P5). We anticipate that only three of them make it in the final PS selection (i.~e., the 0.17\% of the total PS size).

In Eq.~(\ref{eq:selcriterion}), we approximate the 2MASS~$H$ magnitude via the empirical relation:
\begin{equation}
    H = 0.74439+K_s -1.4318\cdot10^{-4}\,T_\mathrm{eff} \textrm { ,}
\end{equation}
calibrated on the tPIC vs.~2MASS cross-match; the coefficients were obtained through an ordinary least-squares fit.

\begin{figure}\centering
    \includegraphics[width=0.99\columnwidth]{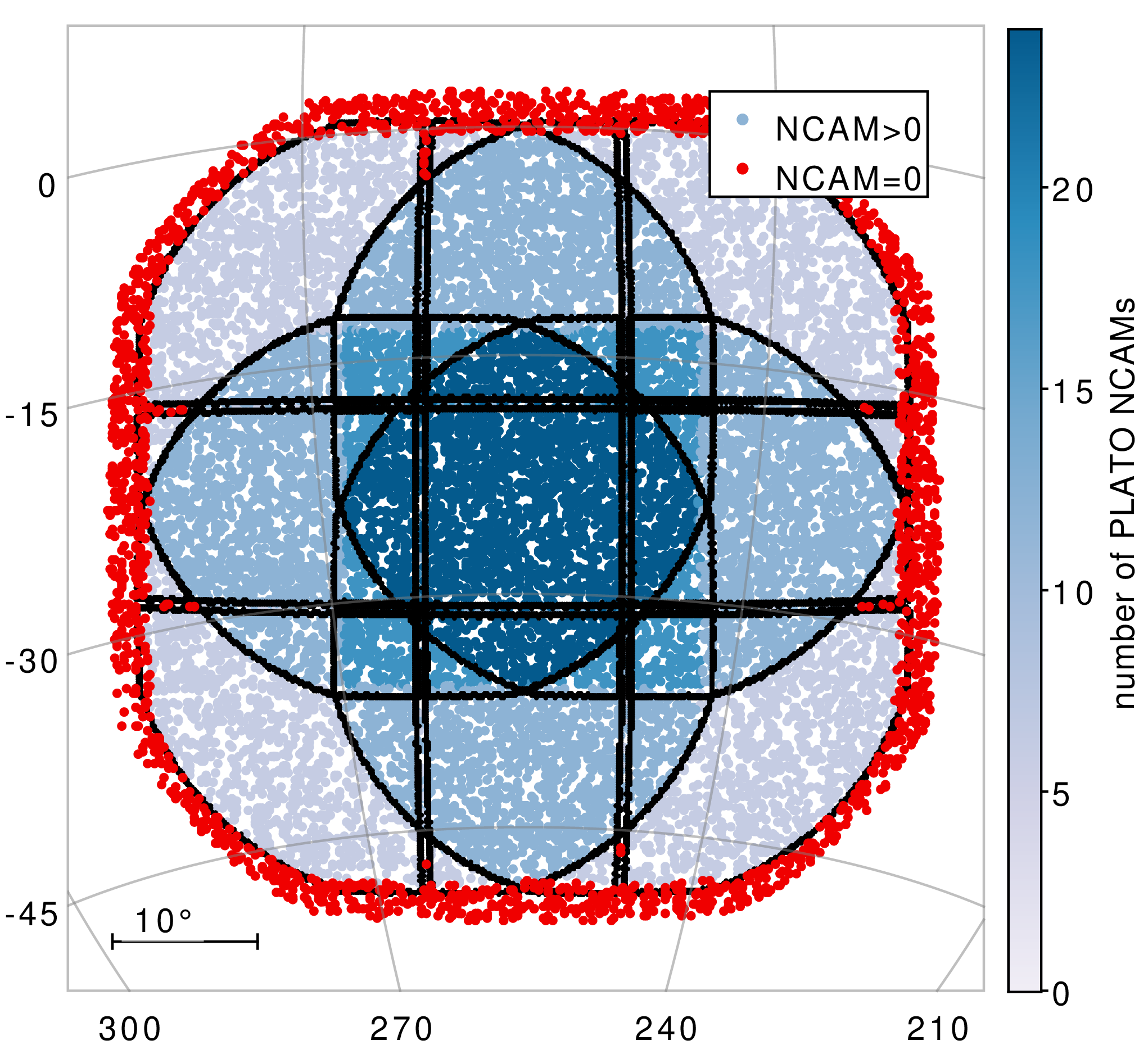}
\caption{Sky map of the LOPS2 field in Galactic coordinates. The $15\,000$ points plotted in blue shades are PS targets falling within the nominal footprint of LOPS2 \citep{Nascimbeni2025} and color-coded according to the nominal number of NCAMs. The $2\,101$ red points are PS stars selected in a safety margin around LOPS2, to take into account alignment and pointing errors (as explained in Sec.~\ref{sec:selection}).}
\label{fig:fov_ncams}
\end{figure}

The output of the selection process is visualized in the radius-magnitude plane in Fig.~\ref{fig:ps_stats}, where the PS stars are color-coded according to their $\mathcal{M}$ and $\mathcal{R}$ values. 
It is worth noting that the total number of PS stars in the tPIC selected by Eq.~(\ref{eq:selcriterion}) is $17\,101$, but only $15\,000$ of them fall within the nominal LOPS2 footprint, for the reasons explained above. This is illustrated in the sky chart plotted in Fig.~\ref{fig:fov_ncams}.

\begin{figure}\centering
    \includegraphics[width=\columnwidth]{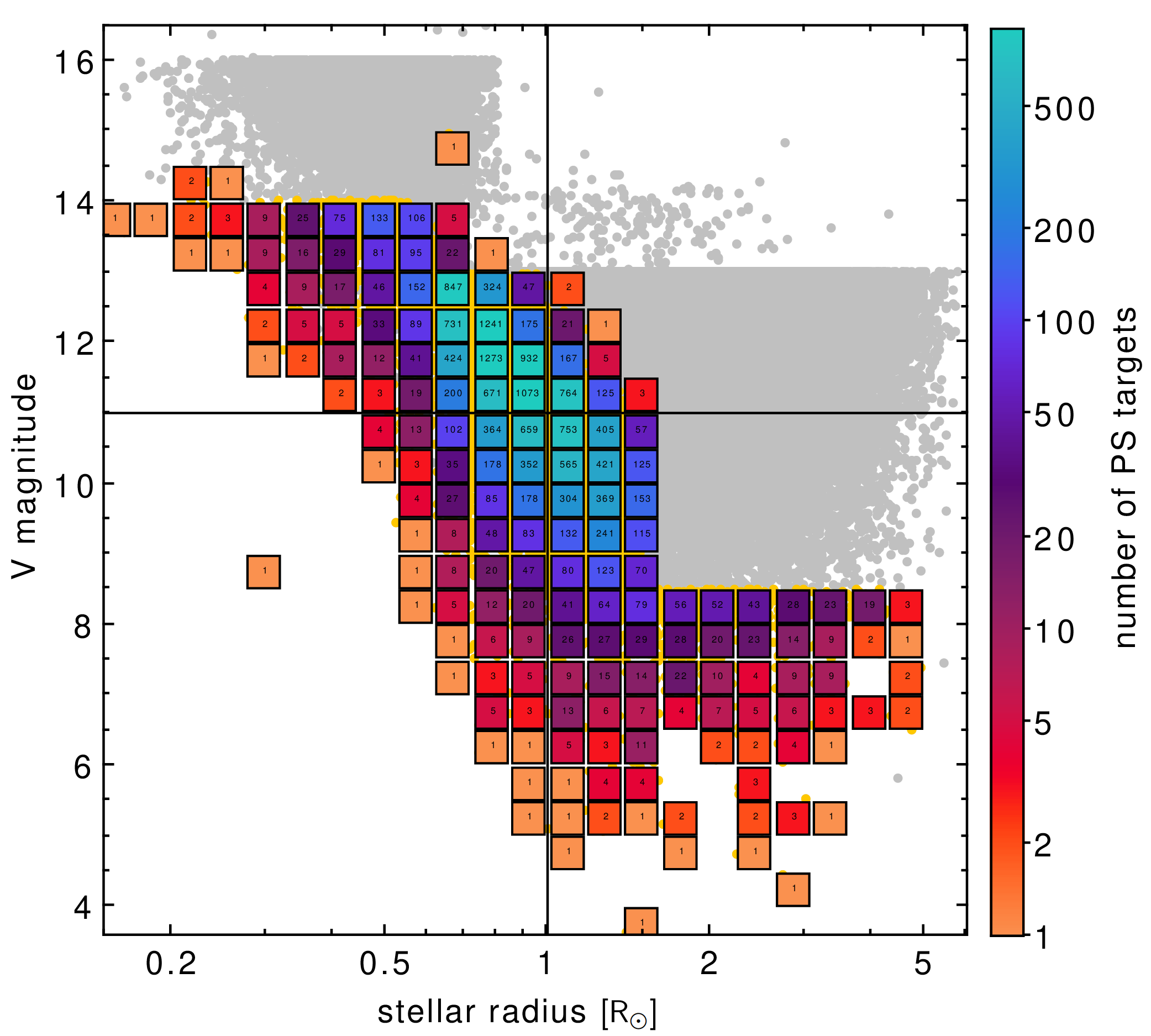}  \vspace{5mm}   \includegraphics[width=\columnwidth]{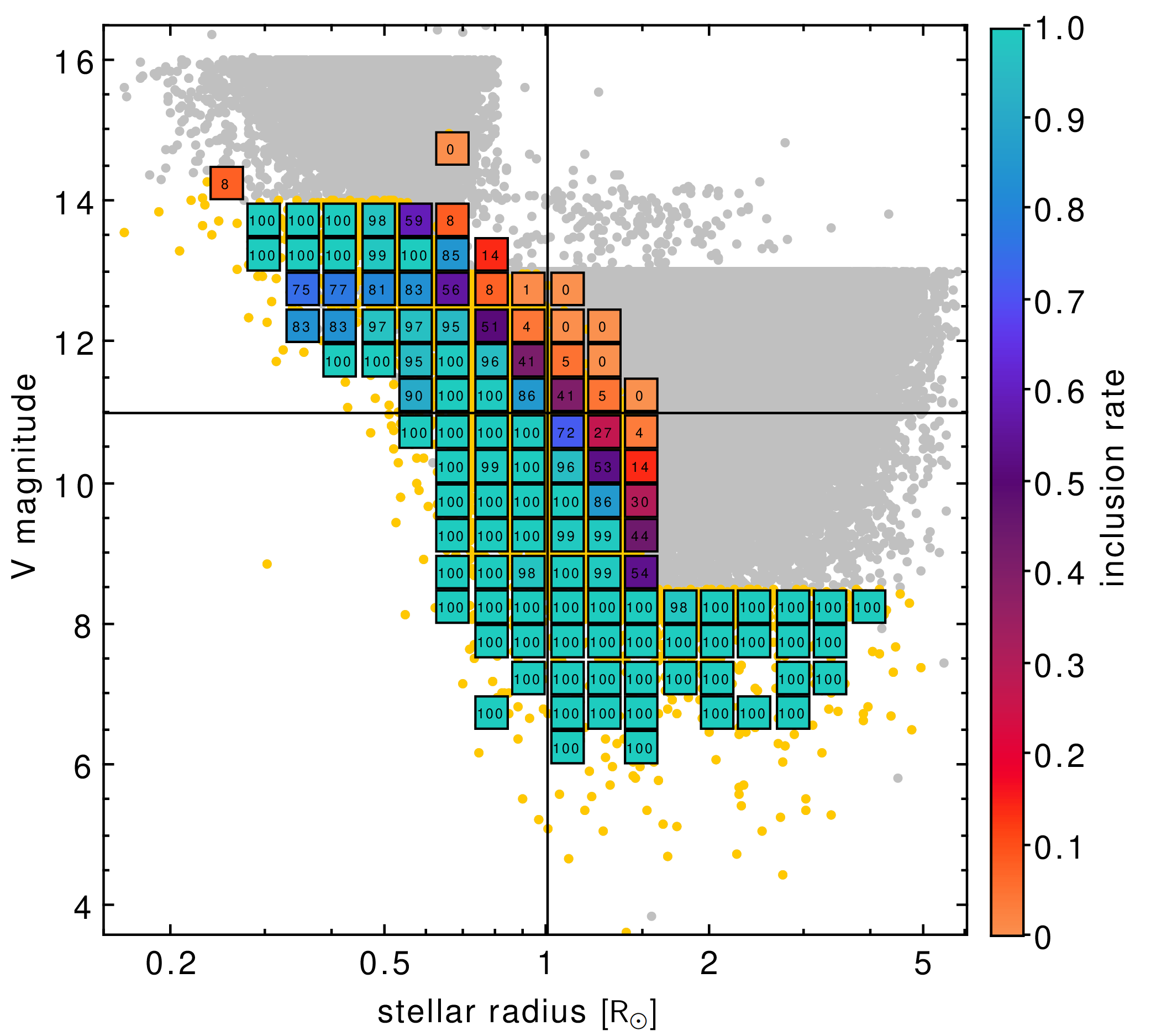}
\caption{Statistical properties of the Prime Sample (Section~\ref{sec:primesample}). \emph{Upper plot:} number of PS stars in each cell of the $V$, $R_\star$ plane. The number is color-coded on a logarithmic scale and labeled on each cell. \emph{Lower plot:} Inclusion rate, i.~e., the fraction of tPIC targets selected in the PS, for each cell of the $V$, $R_\star$ plane. Bright, late-type dwarfs are included with an inclusion rate of 100\% or very close to it, corresponding to the cyan shade of the color scale. Only cells containing at least five tPIC targets are plotted.}
\label{fig:metrics_ps}
\end{figure}

A quick look at the two-dimensional distribution in the same parameter space (Fig.~\ref{fig:metrics_ps}, left panel) reveals that the vast majority of the PS stars are G dwarfs in the $11<V<12$ magnitude range, mostly belonging to the P5 sample. We can also check how the PS selection works by defining the ``inclusion rate'' as the fraction of tPIC stars included in the PS, and computing that number for each cell in the ($V$, $R_\star$) plane (Fig.~\ref{fig:metrics_ps}, right panel). It is evident that our criteria select almost 100\% of bright, late stars: the average inclusion rate is one or very close to one for all stars smaller than 1~$R_\odot$ and brighter than $V=11$, and gradually fades to zero for fainter and/or larger stars.

\begin{figure*}\centering
    \includegraphics[width=0.85\columnwidth]{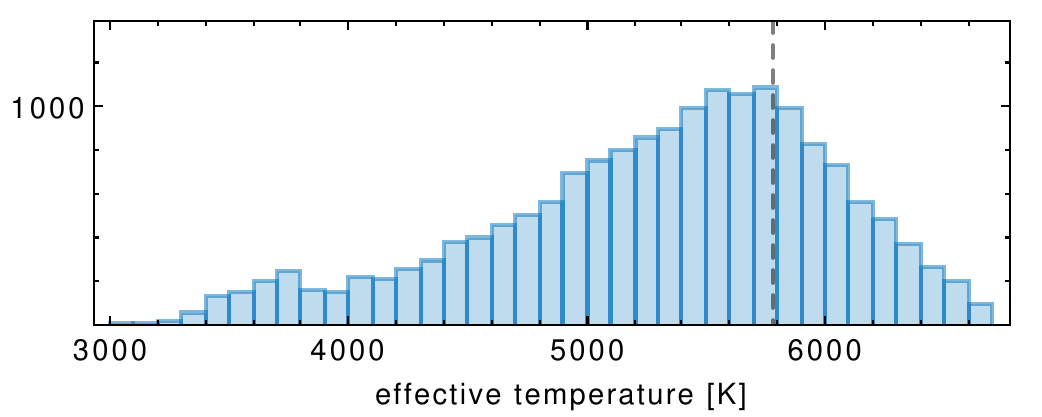}     \includegraphics[width=0.94\columnwidth]{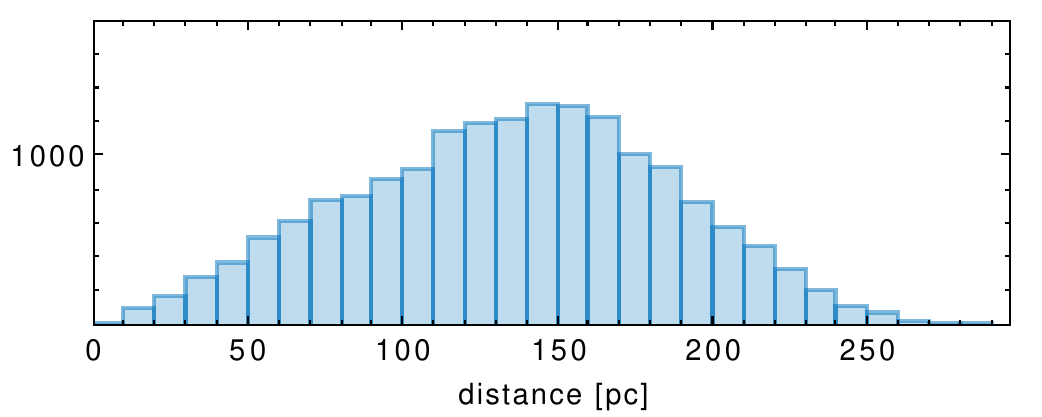} 
    \includegraphics[width=0.85\columnwidth]{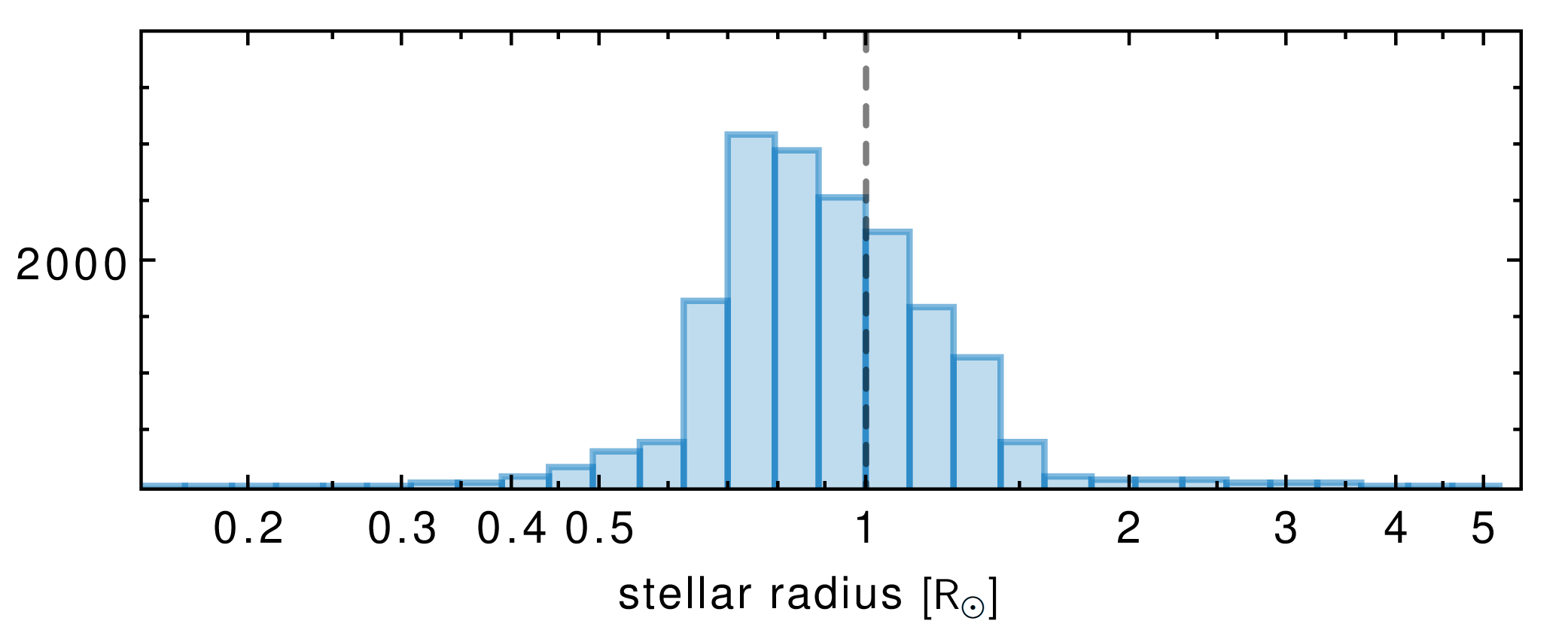}   
    \includegraphics[width=0.85\columnwidth]{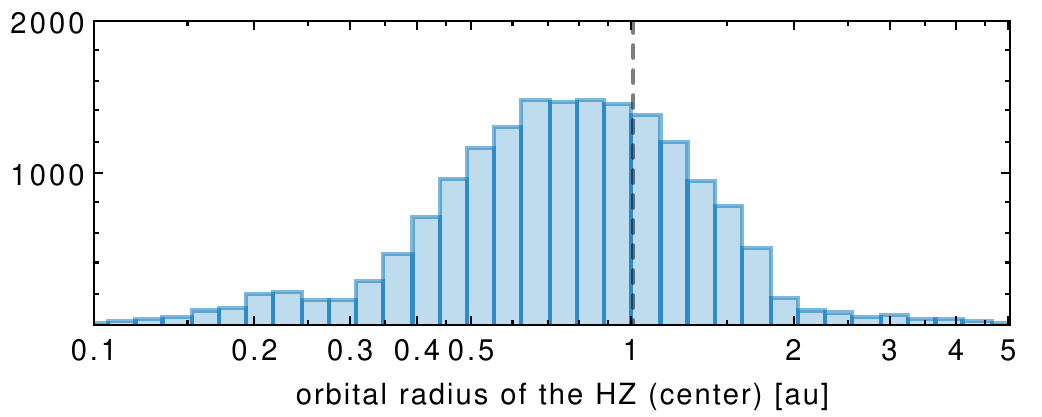}   
    \includegraphics[width=0.85\columnwidth]{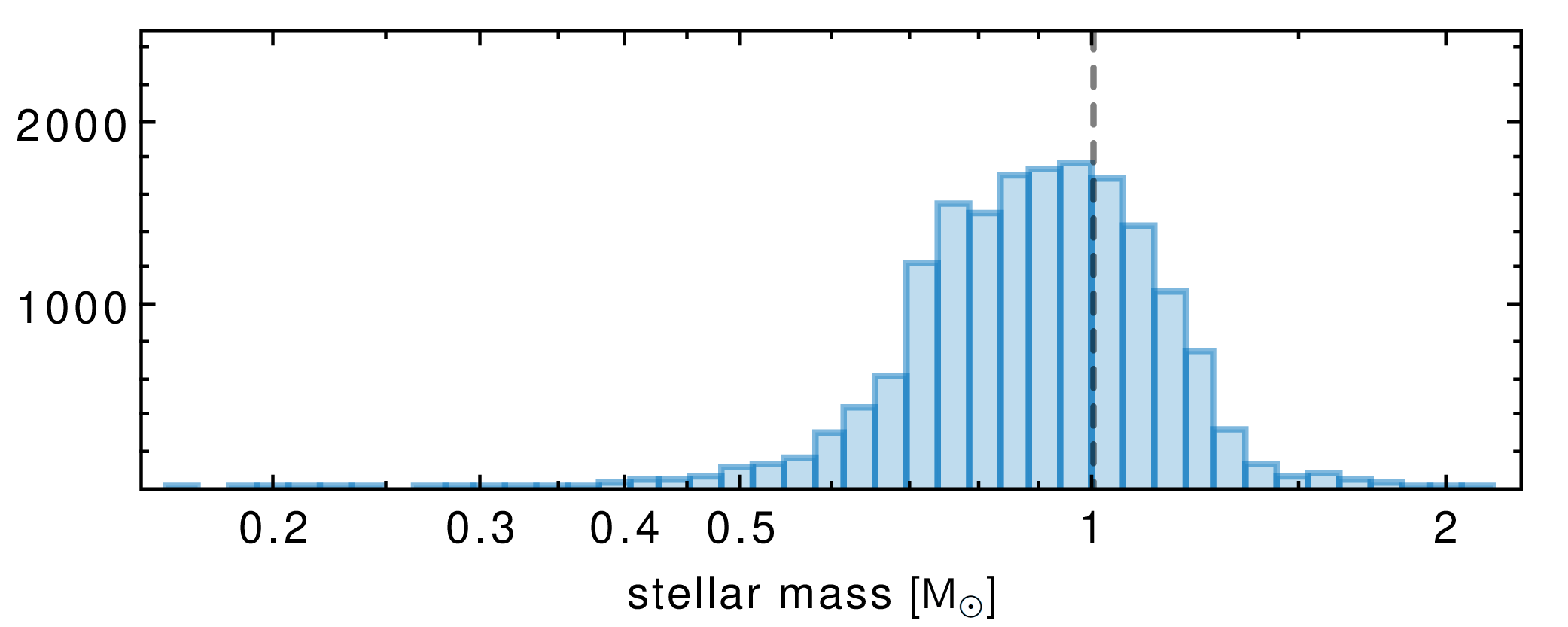} 
    \includegraphics[width=0.85\columnwidth]{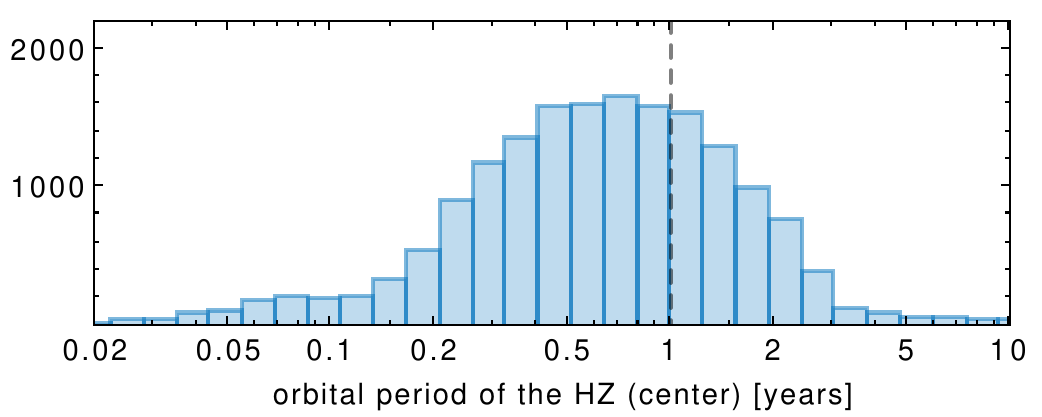}   
\caption{Main parameters of the Prime Sample (Section~\ref{sec:primesample}) stars. \emph{In reading order:} histograms of the distribution in effective temperature $T_\mathrm{eff}$, distance $d$, stellar radius $R_\star$, orbital radius of the HZ (Eq.~\ref{eq:ahz}), stellar mass $M_\star$, orbital period of the HZ (Eq.~\ref{eq:phz}) for all entries flagged as PS stars in the tPIC2.2 ($17\,101$); the values are taken from the same catalog. The dashed vertical lines mark the Solar values: 1~$M_\odot$, 1~$R_\odot$, 5778~K, 1~au, 1~year.}
\label{fig:histograms}
\end{figure*}

\section{Astrophysical properties of the Prime Sample}\label{sec:properties}

The PS we selected is essentially an almost pure sample of solar-type dwarfs, being mostly composed of late F (18\%), G (38\%), and K (37\%) main-sequence stars. The remaining 8\% is made of M dwarfs (due to the inclusion of the bright tail of the P4 subsample) and just a 2\% are slightly evolved subgiants, brighter than $V=8.5$ and belonging to the P2 subsample. Overall, the stellar parameters of the PS are dominated by the G and early-K components: the histograms of the $M_\star$, $R_\star$, $T_\mathrm{eff}$ distribution (Fig.~\ref{fig:histograms}, left panels) are all centered (mode) on Solar values, while the median values (0.90~$M_\odot$, 0.88~$R_\odot$, $5\,420$~K) lie in the range of parameters\footnote{\url{https://github.com/emamajek/SpectralType/blob/master/EEM_dwarf_UBVIJHK_colors_Teff.txt}} for typical G8V-G9V stars \citep{Pecaut2013}. 

\begin{figure*}\centering
    \includegraphics[width=0.9\columnwidth]{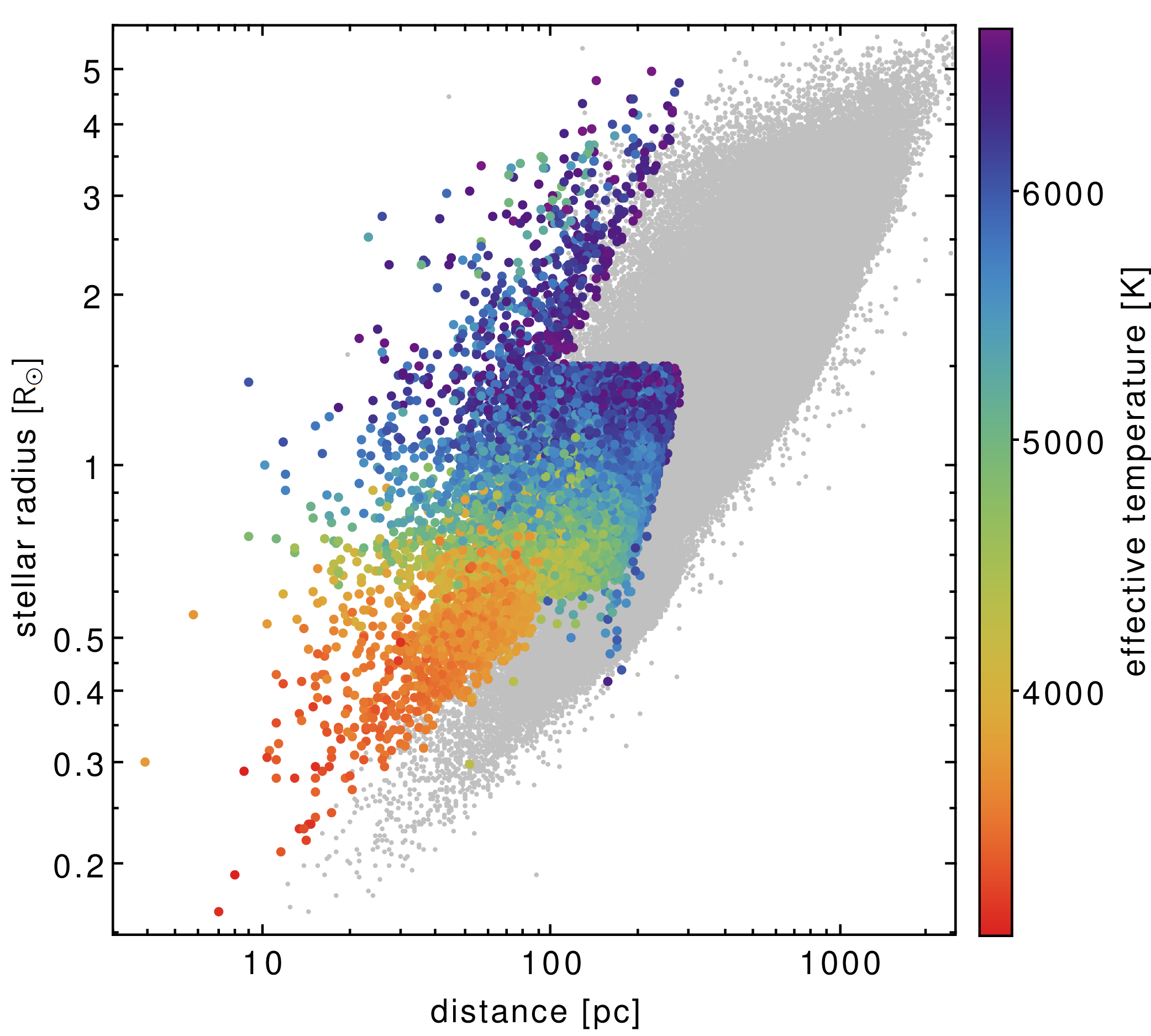}  \hspace{5mm}   \includegraphics[width=0.9\columnwidth]{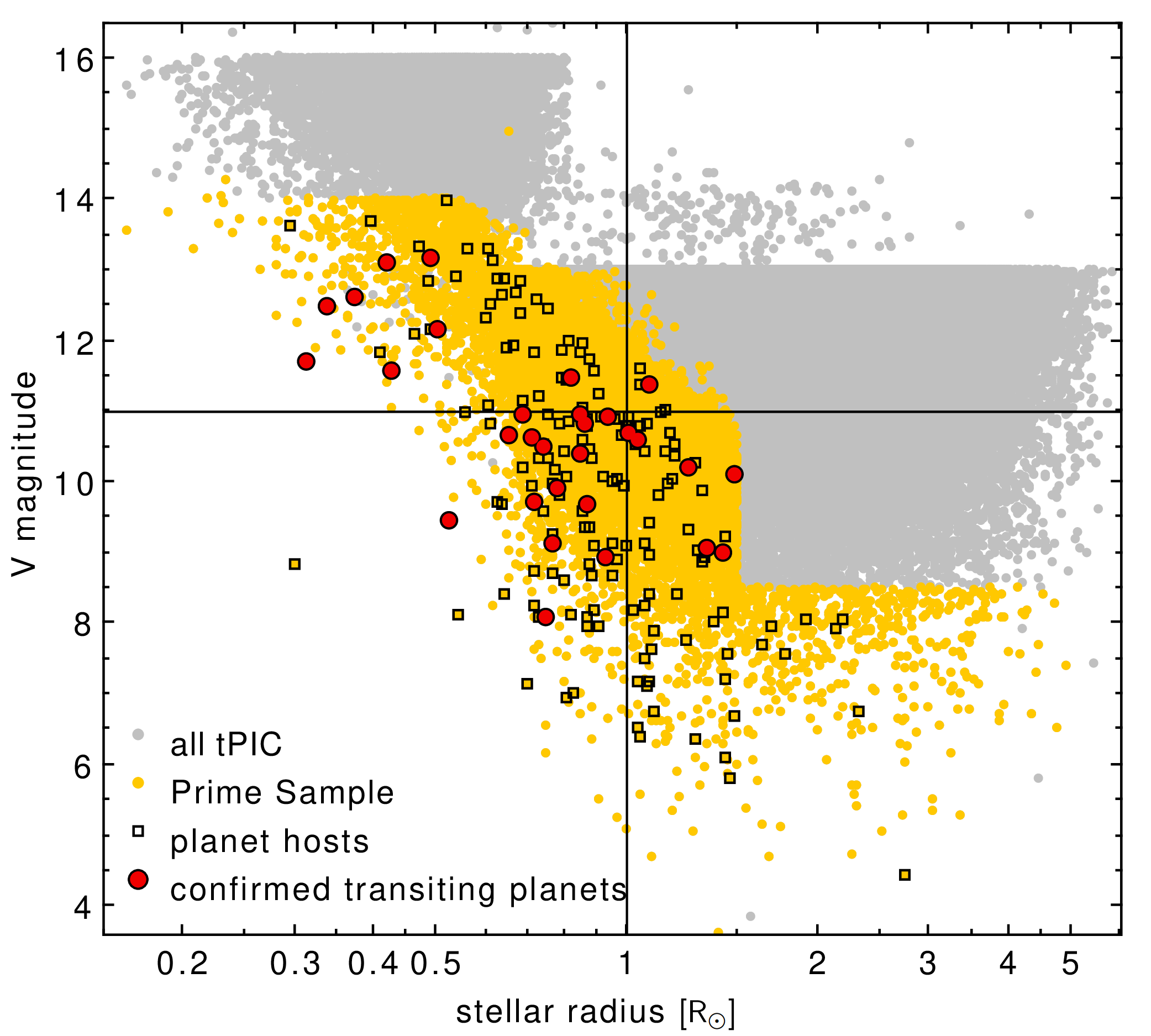}
\caption{Statistical properties of the Prime Sample (Section~\ref{sec:properties}). \emph{Left plot:} stellar radius as a function of distance for the PS stars (circles color-coded according to their effective temperature $T_\mathrm{eff}$). The background light gray points represent the full tPIC sample. \emph{Right plot:} PS stars (yellow points) plotted in the radius-magnitude plane, with known planet hosts marked as open black squares, and hosts of confirmed transiting planets marked with filled red circles. The background light gray points represent the full tPIC sample.}
\label{fig:distance_planets}
\end{figure*}

As expected, PS targets are also on average nearby stars (Fig.~\ref{fig:histograms}, upper right plot): the median distance is 137~pc, with 10th and 90th percentiles at 65 and 200~pc, respectively. The median distance changes significantly for different spectral types (Fig.~\ref{fig:distance_planets}, left panel), going from 50~pc (dM), 120~pc (dK), to 157~pc (dG) and 158~pc (dF). The fact that dG and dF lie at similar distances may appear counterintuitive, but it is due to P2 being forced into the PS, which yields an overabundance of very bright (and relatively nearby) F dwarfs. No star in the PS is more than 300~pc away.

The distribution of the orbital radius and orbital period (respectively) of the HZ for the stars in the PS (middle and lower right panels of Fig.~\ref{fig:histograms})  can be computed using  Eq.~(\ref{eq:ahz}) and Eq.~(\ref{eq:phz}). Noteworthy, both $a_\mathrm{HZ}$ and $P_\mathrm{HZ}$ are centered on values slightly lower than the Earth-Sun reference values of 1~au and 1~year, which implies (as wanted) that the detection of $\geq 3$ transits of HZ planets will be feasible within the nominal mission duration. Note that Eq.~(\ref{eq:ahz}) and Eq.~(\ref{eq:phz}) refer to the HZ as the distance with an equivalent  Earth-Sun irradiance (following \citealt{Kasting1993}), but the actual HZ is rather a range both in $a$ and $P$.

\begin{figure}[t!]\centering
    \includegraphics[width=\columnwidth]{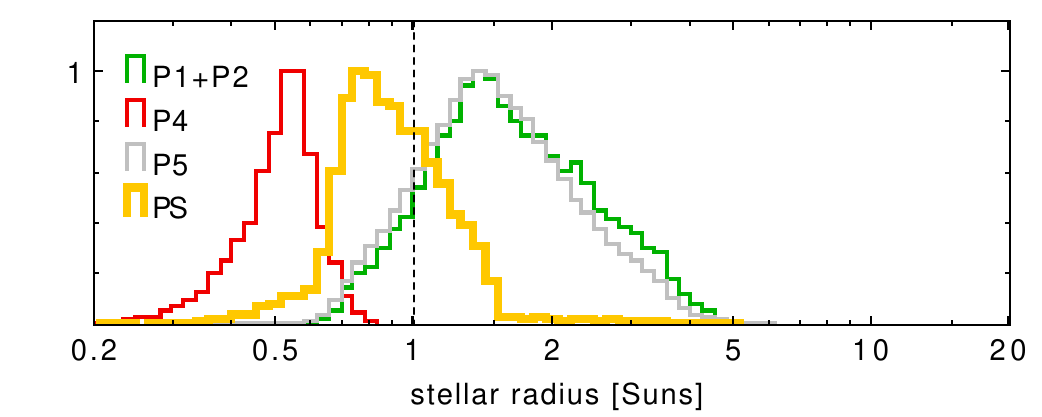}     \includegraphics[width=\columnwidth]{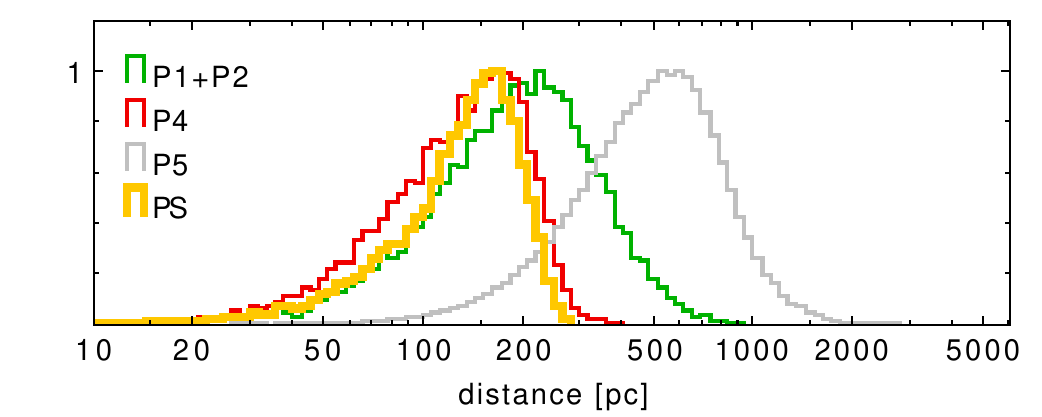} 
    \includegraphics[width=\columnwidth]{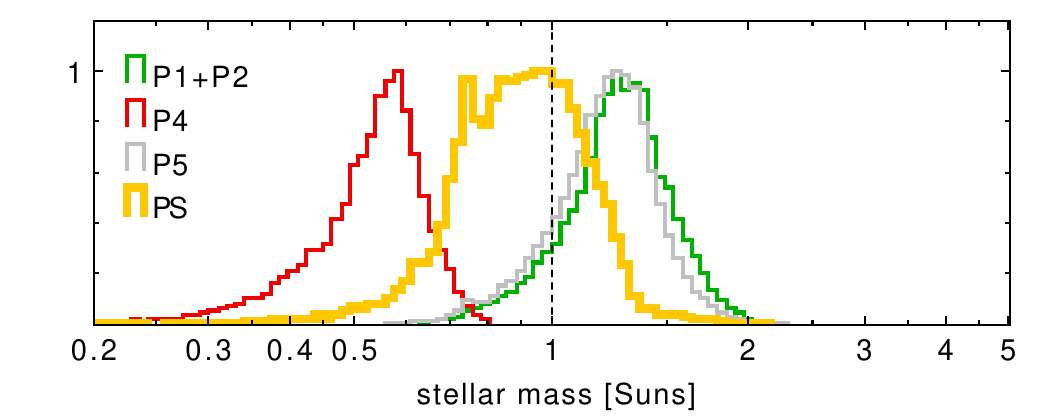}   
    \includegraphics[width=\columnwidth]{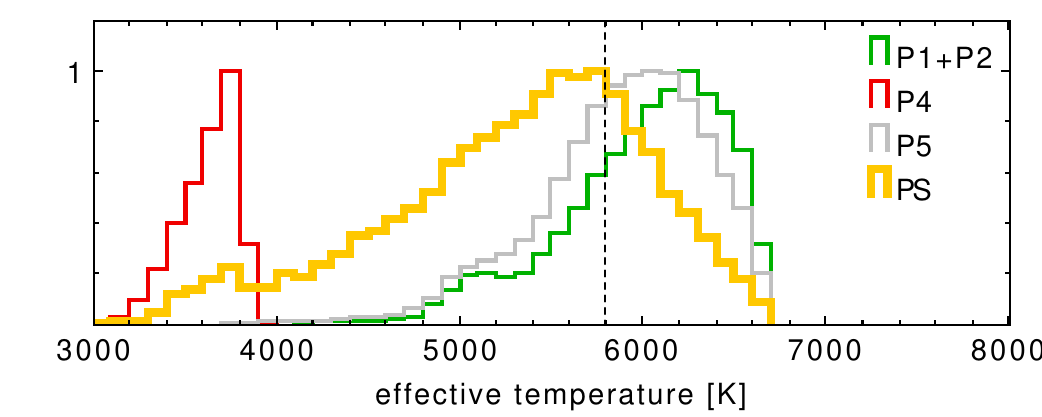}    
\caption{Stellar parameters of the Prime Sample (Section~\ref{sec:primesample}; bold yellow line) stars compared with the PLATO Stellar Samples P1+P2-P4-P5 (Table~\ref{tab:samples}; green, red, gray lines, respectively). All the histograms are normalized to their maximum value. \emph{From top to bottom panel:} histograms of the distribution in stellar radius $R_\star$, distance $d$, stellar mass $M_\star$, and effective temperature $T_\mathrm{eff}$. The dashed vertical lines mark the Solar values: 1~$M_\odot$, 1~$R_\odot$, 5778~K.}
\label{fig:histograms_p}
\end{figure}

As for the primary goal of detecting and confirming a true Earth analogue, the most valuable part of the PS is its bright tail, in particular for targets with $V<11$, for which  high-precision RVs are not excessively expensive and the follow-up is facilitated by the availability of accurate stellar parameters from asteroseismic analysis. About $7\,000$ targets (more than 40\% of the PS) are $V<11$ FGK dwarfs. We refer the reader to Table~\ref{tab:psnumbers} for a more comprehensive description of the different PS subsets, and to the histograms of Fig.~\ref{fig:histograms_p} for a more direct comparison between the PS and the main PLATO Stellar Samples (P1-P2-P4-P5, as defined in Table~\ref{tab:samples}).

\begin{table}\centering\small
\caption{Subsets of the PLATO Prime Sample.}
\begin{tabular}{lccp{3.5cm}}
\hline\hline
subset & number & fraction & definition/comment \\ \hline \noalign{\smallskip}
P1 & 5516 & 32\%& dFGK+sG, $V<11$, $\textrm{NSR}<50$~ppm (Table~\ref{tab:samples})\\
``pure'' P1 & 4648 & 27\% & dFGK+sG, $\textrm{NSR}<50$~ppm (Table~\ref{tab:samples}), excluding P2 \\
P2 & 868 & 5\% & dFGK+sG, $V<8.5$, $\textrm{NSR}<50$~ppm (Table~\ref{tab:samples}) \\
P4 & 863 & 5\% & dM, $V<16$ \citep{Prisinzano2026}\\
P5 & 10720 & 63\% &  dFGK+sG, $V<13$, not in P1 or P2 (Table~\ref{tab:samples})\\
``bright'' & 7005 & 41\% & $V<11$ \\
dF & 3058 & 18\% & F dwarfs (as def.~in \citealt{Nascimbeni2022}) \\
dG & 6527 & 38\% & G dwarfs  (as def.~in \citealt{Nascimbeni2022}) \\
dK & 6311 & 37\% & K dwarfs  (as def.~in \citealt{Nascimbeni2022})  \\
subgiants & 342 & 2\% & subgiants (as def.~in \citealt{Nascimbeni2022})  \\
bright dF & 2732 & \dots & $V<11$ F dwarfs \\
bright dG & 3123 & \dots & $V<11$ G dwarfs \\
bright dK & 811 & \dots & $V<11$ K dwarfs \\
planet hosts & 184 & \dots & all types (incl. candidates)\\
planet hosts & 84 & \dots & all types (confirmed)\\
planet hosts & 30 & \dots & transiting, confirmed\\
NCAM6 & 4268 & 25\% & observed with 6 NCAMs\\
NCAM12 & 6486 & 38\% & observed with 12 NCAMs\\
NCAM18 & 1314 & 8\% & observed with 18 NCAMs\\
NCAM24 & 2932 & 17\% & observed with 24 NCAMs\\
NCAM0 & 2101 & 12\% & outside the nominal LOPS2 field\\ \hline
\end{tabular}\label{tab:psnumbers}
\tablefoot{The columns give: the name of the subset, the number of targets in the subset, the fraction with respect to the total of 17101 targets in the ``full'' PS (including stars outside the nominal LOPS2 field), and how the subset is defined.}
\end{table}

\label{sec:additional}

Planet-hosting stars (list updated on 8 April 2025; next update before the beginning of science observations) are not forced into the PS, but 184 among them meet the PS criteria (out of the 789 stars in the full tPIC,  including non-transiting and/or candidate planets). Of these systems, 84 contain at least one confirmed planet, and 30 are transiting systems as well (Fig.~\ref{fig:distance_planets}, right panel). High-multiplicity systems such as TOI-700 ($N=4$; \citealt{Gilbert2020,Gilbert2023}) and HD~23472 ($N=5$; \citealt{Trifonov2019,Teske2021}) are included in the PS, as well as other systems of high scientific interest, such as TOI-712, TOI-451, TOI-201, TOI-431, L98-59, for which we refer the reader to a more detailed discussion in paper II \citep{Nascimbeni2025} and in \citet{Eschen2024}.

PS stars are distributed uniformly over the LOPS2 field with an average density of about 6~$\textrm{deg}^{-2}$. Not surprisingly, their observability from the Northern Hemisphere is limited, but not zero. Among the $15\,000$ PS stars located in the nominal field, 343 (2.3\%) are located northern of declination $\delta = -25^\circ$, 1568 (10.4\%) northern of $\delta = -30^\circ$, and 3206 (21.4\%) northern of $\delta = -35^\circ$, opening the opportunity of follow-up from moderate-latitude observatories such as Mauna Kea ($\phi\simeq 20^\circ$) and La Palma ($\phi\simeq 28.5^\circ$) for a non-negligible fraction of targets. 

\section{4MOST preparatory observations}\label{sec:4most}

The PLATO southern field, including the PS, will also be observed as part of the 4MOST high-resolution disc and bulge survey, 4MIDABLE-HR (\citealt{Bensby2019}; Bergemann et al., in prep.). 4MOST stands for Multi-Object Spectroscopic Telescope, and its operations and scheduling are described by \citet{deJong2019} and \citet{Walcher2019}. 4MOST had its first light in October 2025 and the start of 4MOST survey operations is planned for June 2026 and the survey operations will continue during five years. With a 4~$\mathrm{deg}^2$ FOV and high multiplexing capability, the telescope will be able to observe up to 800 targets at high resolution ($R \approx 20\,000$)  and  up to 1600 targets at low resolution ($R \approx 5\,000$), although the resolving power depends slightly on wavelength. Specifically, this sub-sample will be part of the so-called S4:10 sub-survey, and it will be observed by 4MOST as part of the 4MOST-PLATO Memorandum of Understanding (MoU). The entire 4MIDABLE-HR S4:10 dataset contains $192\,609$ stars that includes P1, P2, P4 and P5 samples in the PLATO southern field.

The 4MIDABLE-HR survey will obtain high-resolution spectra of the PS, aiming for a S/N of at least 100 per \AA{} in the green. The spectra will be taken simultaneously in three spectral windows: $3926$--$4355$~\AA{}, $5160$--$5730$~\AA{}, and $6100$--$6790$~\AA{}, and multi-epoch observations may be available for selected stars. These broad-band spectra cover the main diagnostic lines of the following chemical elements: Li, CNO, $\alpha$ elements, including Mg, Si, and Ca, odd-Z elements (Na, Al), Fe-peak elements including Ti, Mn, Co, and Ni, as well as slow- and rapid-neutron capture process elements (Sr, Ba, Y, Eu). Due to the coverage in the near-UV, these spectra will also enable analyses of stellar chromospheric activity. These measurements can be performed using the \ion{Ca}{II} H \& K and H$_{\alpha}$ lines. In addition, surface projected rotation velocities ($v \sin i$, with $i$ the inclination) will be measured with characteristic uncertainties of several km s$^{-1}$.

\section{Discussion and conclusions}\label{conclusions}

Throughout this work, we have described how we selected from the tPIC2.2  the PLATO PS for the first LOP observing field (LOPS2). This subset of the tPIC will be flagged in PIC2.2 and made public to the astronomical community at the opening of the first PLATO GO call for proposals\footnote{The PLATO GO call and the PLATO Archive will be published at this URLs: \url{https://pax.esac.esa.int/plato}, \url{https://www.cosmos.esa.int/web/plato/ao-1}, respectively}. PS stars cannot be proposed as targets of any GO proposal, regardless of the underlying science case. Updates to the PS list will be possible in the future, for instance, in case of future GO calls, or additional LOP fields to be pointed. 

Two crucial ingredients of our final selection criteria (summarized by Eq.~\ref{eq:selcriterion}) are the metrics $\mathcal{M}$ (Eq.~\ref{eq:m}) and $\mathcal{R}$ (Eq.~\ref{eq:r}) developed for this purpose and linked to the S/N expected for a HZ Earth analogue in order to maximize: 1) the photometric detection probability by PLATO, and 2) the maximum capability of RV follow-up by ground-based optical spectrographs. We have shown how both metrics can be fully generalized and applied to any other observing program focused on the detection and characterization of transiting, habitable planets.
We also emphasize how, in principle, one could easily adapt our selection criteria (\ref{eq:selcriterion}) to a different assumption on the reference semi-major axis and radius of HZ planets. With a more ``optimistic'' definition of $R_\mathrm{p}=1.5$~$R_\oplus$ and $a_\mathrm{HZ}=0.75$~au (corresponding to the inner edge of the HZ as defined by \citealt{Kasting1993}), the $\mathcal{M}$ metric would increase by a factor of $(R_\mathrm{p}/R_\oplus)^2(a_\mathrm{HZ}/\mathrm{au})^{-1/2}\simeq 2.6$. The change in the normalization factor of the $\mathcal{R}$ metric, on the other hand, would have no impact on the selection, since the size of the PS is fixed and  $\mathcal{R}$ acts as a free parameter to be adjusted in order to get $15\, 000$ targets. Such a modified PS, for instance, would still contain 85\% of the targets in common with the PIC2.2 PS.

The $\mathcal{M}$ metric is also currently the basis to compute the tPIC target ranking, as listed in the  \texttt{tPICScientificRanking} column of PIC2.2. This number will be essential to rank all the available targets according to their scientific priority, allowing the PLATO Target Programming Tool (TPT) to make choices when, for technical limitations, two targets cannot be scheduled at the same time. The \texttt{tPICScientificRanking} is defined as the $\mathcal{M}$ metric normalized to one, with the only exception of planet-hosting stars (including hosts of candidate planets) for which the number is forced to one, i.~e., to the maximum priority.

As already discussed in detail in \citet{Rauer2025}, ESA will issue the first Guest Observer (GO) call (current release date is 7 April 2026) with the aim to allow for the use of the PLATO instruments by the worldwide community. A total of 8\% of PLATO's overall telemetry budget is available for selected competitive proposals covering science cases outside of PLATO's core science program. For proposals to be eligible for this first call, the suggested targets must be situated in LOPS2. They cannot be part of the PS defined in this paper. For more details, we refer to the actual future ESA call.

\begin{acknowledgements}
This work presents results from the European Space Agency (ESA) space mission PLATO. The PLATO payload, the PLATO Ground Segment and PLATO data processing are joint developments of ESA and the PLATO mission consortium (PMC). Funding for the PMC is provided at national levels, in particular by countries participating in the PLATO Multilateral Agreement (Austria, Belgium, Czech Republic, Denmark, France, Germany, Italy, Netherlands, Portugal, Spain, Sweden, Switzerland, Norway, and United Kingdom) and institutions from Brazil. Members of the PLATO Consortium can be found at \url{https://platomission.com/}. The ESA PLATO mission website is \url{https://www.cosmos.esa.int/plato}. We thank the teams working for PLATO for all their work.\\
This work has made use of data from the European Space Agency (ESA) mission {\it Gaia} (\url{https://www.cosmos.esa.int/gaia}), processed by the {\it Gaia} Data Processing and Analysis Consortium (DPAC,
\url{https://www.cosmos.esa.int/web/gaia/dpac/consortium}). Funding for the DPAC has been provided by national institutions, in particular the institutions participating in the {\it Gaia} Multilateral Agreement. \\
VN, GP, MM, SB, SD, VG, LM, IP, LP, RR, DN acknowledge support from PLATO ASI-INAF agreements n. 2022-28-HH.0. GA, PM and SM acknowledge financial support from PLATO ASI-INAF agreements n.~2022-14-HH.0 and n.~2025-33.HH.0.  
JMMH is funded by Spanish MICIU/AEI/10.13039/501100011033 and ERDF/EU grant PID2023-147338NB-C21.
LP acknowledges support from the Italian Ministero dell'Università e della Ricerca and
the European Union - Next Generation EU through project PRIN 2022 PM4JLH ``Know your little neighbours: characterising low-mass stars and planets in the Solar neighbourhood''. CA acknowledges financial support from the BELgian federal Science Policy Office (BELSPO) through the PRODEX grant for PLATO. 
MB is supported from the European Research Council (ERC) under the European Union’s Horizon 2020 research and innovation programme (Grant agreement No. 949173).\\

We thank M.~Lund, L.~Mancini, D.~Godoy-Rivera, M.~G.~Albanese and the anonymous referee for the useful comments and suggestions.
\end{acknowledgements}

\bibliographystyle{aa}
\bibliography{biblio}

\begin{thebibliography}{40}
\expandafter\ifx\csname natexlab\endcsname\relax\def\natexlab#1{#1}\fi

\bibitem[{{Auvergne} {et~al.}(2009){Auvergne}, {Bodin}, {Boisnard}, {Buey}, {Chaintreuil}, {Epstein}, {Jouret}, {Lam-Trong}, {Levacher}, {Magnan}, {Perez}, {Plasson}, {Plesseria}, {Peter}, {Steller}, {Tiph{\`e}ne}, {Baglin}, {Agogu{\'e}}, {Appourchaux}, {Barbet}, {Beaufort}, {Bellenger}, {Berlin}, {Bernardi}, {Blouin}, {Boumier}, {Bonneau}, {Briet}, {Butler}, {Cautain}, {Chiavassa}, {Costes}, {Cuvilho}, {Cunha-Parro}, {de Oliveira Fialho}, {Decaudin}, {Defise}, {Djalal}, {Docclo}, {Drummond}, {Dupuis}, {Exil}, {Faur{\'e}}, {Gaboriaud}, {Gamet}, {Gavalda}, {Grolleau}, {Gueguen}, {Guivarc'h}, {Guterman}, {Hasiba}, {Huntzinger}, {Hustaix}, {Imbert}, {Jeanville}, {Johlander}, {Jorda}, {Journoud}, {Karioty}, {Kerjean}, {Lafond}, {Lapeyrere}, {Landiech}, {Larqu{\'e}}, {Laudet}, {Le Merrer}, {Leporati}, {Leruyet}, {Levieuge}, {Llebaria}, {Martin}, {Mazy}, {Mesnager}, {Michel}, {Moalic}, {Monjoin}, {Naudet}, {Neukirchner}, {Nguyen-Kim}, {Ollivier}, {Orcesi}, {Ottacher}, {Oulali}, {Parisot}, {Perruchot}, {Piacentino},
  {Pinheiro da Silva}, {Platzer}, {Pontet}, {Pradines}, {Quentin}, {Rohbeck}, {Rolland}, {Rollenhagen}, {Romagnan}, {Russ}, {Samadi}, {Schmidt}, {Schwartz}, {Sebbag}, {Smit}, {Sunter}, {Tello}, {Toulouse}, {Ulmer}, {Vandermarcq}, {Vergnault}, {Wallner}, {Waultier}, \& {Zanatta}}]{Auvergne2009}
{Auvergne}, M., {Bodin}, P., {Boisnard}, L., {et~al.} 2009, \aap, 506, 411

\bibitem[{{Bensby} {et~al.}(2019){Bensby}, {Bergemann}, {Rybizki}, {Lemasle}, {Howes}, {Kovalev}, {Agertz}, {Asplund}, {Barklem}, {Battistini}, {Casagrande}, {Chiappini}, {Church}, {Feltzing}, {Ford}, {Gerhard}, {Kushniruk}, {Kordopatis}, {Lind}, {Minchev}, {McMillan}, {Rix}, {Ryde}, \& {Traven}}]{Bensby2019}
{Bensby}, T., {Bergemann}, M., {Rybizki}, J., {et~al.} 2019, The Messenger, 175, 35

\bibitem[{{B{\"o}rner} {et~al.}(2024){B{\"o}rner}, {Paproth}, {Cabrera}, {Pertenais}, {Rauer}, {Mas-Hesse}, {Pagano}, {Alvarez}, {Erikson}, {Grie{\ss}bach}, {Levillain}, {Magrin}, {Mogulsky}, {Niemi}, {Prod'homme}, {Regibo}, {De Ridder}, {Rockstein}, {Samadi}, {Serrano-Velarde}, {Smith}, {Verhoeve}, \& {Walton}}]{Borner2024}
{B{\"o}rner}, A., {Paproth}, C., {Cabrera}, J., {et~al.} 2024, Experimental Astronomy, 58, 1

\bibitem[{{Borucki} {et~al.}(2010){Borucki}, {Koch}, {Basri}, {Batalha}, {Brown}, {Caldwell}, {Caldwell}, {Christensen-Dalsgaard}, {Cochran}, {DeVore}, {Dunham}, {Dupree}, {Gautier}, {Geary}, {Gilliland}, {Gould}, {Howell}, {Jenkins}, {Kondo}, {Latham}, {Marcy}, {Meibom}, {Kjeldsen}, {Lissauer}, {Monet}, {Morrison}, {Sasselov}, {Tarter}, {Boss}, {Brownlee}, {Owen}, {Buzasi}, {Charbonneau}, {Doyle}, {Fortney}, {Ford}, {Holman}, {Seager}, {Steffen}, {Welsh}, {Rowe}, {Anderson}, {Buchhave}, {Ciardi}, {Walkowicz}, {Sherry}, {Horch}, {Isaacson}, {Everett}, {Fischer}, {Torres}, {Johnson}, {Endl}, {MacQueen}, {Bryson}, {Dotson}, {Haas}, {Kolodziejczak}, {Van Cleve}, {Chandrasekaran}, {Twicken}, {Quintana}, {Clarke}, {Allen}, {Li}, {Wu}, {Tenenbaum}, {Verner}, {Bruhweiler}, {Barnes}, \& {Prsa}}]{Borucki2010}
{Borucki}, W.~J., {Koch}, D., {Basri}, G., {et~al.} 2010, Science, 327, 977

\bibitem[{{Bouchy} {et~al.}(2025){Bouchy}, {Doyon}, {Pepe}, {Melo}, {Artigau}, {Malo}, {Wildi}, {Baron}, {Delfosse}, {De Medeiros}, {Rebolo}, {Santos}, {Wade}, {Allart}, {Al Moulla}, {Blind}, {Cadieux}, {Canto Martins}, {Cook}, {Dumusque}, {Frensch}, {Genest}, {Gonz{\'a}lez Hern{\'a}ndez}, {Grieves}, {Lo Curto}, {Lovis}, {Mignon}, {Nielsen}, {Poulin-Girard}, {Rasilla}, {Reshetov}, {Sosnowska}, {Sordet}, {Saint-Antoine}, {Su{\'a}rez Mascare{\~n}o}, {Thibault}, {Vall{\'e}e}, {Vandal}, {Abreu}, {Aguiar}, {Allain}, {Arial}, {Auger}, {Barros}, {Bazinet}, {Benneke}, {Bonfils}, {Boucher}, {Bourrier}, {Bovay}, {Broeg}, {Brousseau}, {Bruniquel}, {Bryan}, {Cabral}, {Carmona}, {Carteret}, {Challita}, {Chazelas}, {Cloutier}, {Coelho}, {Cointepas}, {Conod}, {Cowan}, {Cristo}, {Gomes da Silva}, {Dauplaise}, {Darveau-Bernier}, {de Lima Gomes}, {de Freitas}, {Delgado-Mena}, {Delisle}, {Ehrenreich}, {Faria}, {Figueira}, {Fontinele}, {Forveille}, {Gagn{\'e}}, {Genolet}, {T{\'e}mich}, {Hernandez}, {Hobson}, {Hoeijmakers},
  {Hubin}, {Jahandar}, {Jayawardhana}, {K{\"a}ufl}, {Kerley}, {Kolb}, {Krishnamurthy}, {Lafreni{\`e}re}, {Lamontagne}, {Larue}, {Leath}, {L'Heureux}, {de Castro Le{\~a}o}, {Lim}, {Martins}, {Matthews}, {Mayer}, {Messias}, {Metchev}, {Moranta}, {Mordasini}, {Mounzer}, {Nari}, {Osborn}, {Ouellet}, {Otegi}, {Parc}, {Pasquini}, {Passegger}, {Pelletier}, {Peroux}, {Piaulet-Ghorayeb}, {Plotnykov}, {Pompei}, {Rowe}, {Sarajlic}, {Segovia}, {Seidel}, {S{\'e}gransan}, {Schnell}, {Costa Silva}, {Srivastava}, {Stefanov}, {Teixeira}, {Udry}, {Valencia}, {Vaulato}, {Wardenier}, {Wehbe}, {Weisserman}, {Wevers}, {Yariv}, \& {Zins}}]{Bouchy2025}
{Bouchy}, F., {Doyon}, R., {Pepe}, F., {et~al.} 2025, \aap, 700, A10

\bibitem[{{Cayrel de Strobel}(1996)}]{deStrobel1996}
{Cayrel de Strobel}, G. 1996, \aapr, 7, 243

\bibitem[{{Csizmadia} {et~al.}(2013){Csizmadia}, {Pasternacki}, {Dreyer}, {Cabrera}, {Erikson}, \& {Rauer}}]{Czismadia2013}
{Csizmadia}, S., {Pasternacki}, T., {Dreyer}, C., {et~al.} 2013, \aap, 549, A9

\bibitem[{{Csizmadia} {et~al.}(2023){Csizmadia}, {Smith}, {K{\'a}lm{\'a}n}, {Cabrera}, {Klagyivik}, {Chaushev}, \& {Lam}}]{Csizmadia2023}
{Csizmadia}, S., {Smith}, A.~M.~S., {K{\'a}lm{\'a}n}, S., {et~al.} 2023, \aap, 675, A106

\bibitem[{{de Jong} {et~al.}(2019){de Jong}, {Agertz}, {Berbel}, {Aird}, {Alexander}, {Amarsi}, {Anders}, {Andrae}, {Ansarinejad}, {Ansorge}, {Antilogus}, {Anwand-Heerwart}, {Arentsen}, {Arnadottir}, {Asplund}, {Auger}, {Azais}, {Baade}, {Baker}, {Baker}, {Balbinot}, {Baldry}, {Banerji}, {Barden}, {Barklem}, {Barth{\'e}l{\'e}my-Mazot}, {Battistini}, {Bauer}, {Bell}, {Bellido-Tirado}, {Bellstedt}, {Belokurov}, {Bensby}, {Bergemann}, {Bestenlehner}, {Bielby}, {Bilicki}, {Blake}, {Bland-Hawthorn}, {Boeche}, {Boland}, {Boller}, {Bongard}, {Bongiorno}, {Bonifacio}, {Boudon}, {Brooks}, {Brown}, {Brown}, {Br{\"u}ggen}, {Brynnel}, {Brzeski}, {Buchert}, {Buschkamp}, {Caffau}, {Caillier}, {Carrick}, {Casagrande}, {Case}, {Casey}, {Cesarini}, {Cescutti}, {Chapuis}, {Chiappini}, {Childress}, {Christlieb}, {Church}, {Cioni}, {Cluver}, {Colless}, {Collett}, {Comparat}, {Cooper}, {Couch}, {Courbin}, {Croom}, {Croton}, {Daguis{\'e}}, {Dalton}, {Davies}, {Davis}, {de Laverny}, {Deason}, {Dionies}, {Disseau}, {Doel},
  {D{\"o}scher}, {Driver}, {Dwelly}, {Eckert}, {Edge}, {Edvardsson}, {Youssoufi}, {Elhaddad}, {Enke}, {Erfanianfar}, {Farrell}, {Fechner}, {Feiz}, {Feltzing}, {Ferreras}, {Feuerstein}, {Feuillet}, {Finoguenov}, {Ford}, {Fotopoulou}, {Fouesneau}, {Frenk}, {Frey}, {Gaessler}, {Geier}, {Gentile Fusillo}, {Gerhard}, {Giannantonio}, {Giannone}, {Gibson}, {Gillingham}, {Gonz{\'a}lez-Fern{\'a}ndez}, {Gonzalez-Solares}, {Gottloeber}, {Gould}, {Grebel}, {Gueguen}, {Guiglion}, {Haehnelt}, {Hahn}, {Hansen}, {Hartman}, {Hauptner}, {Hawkins}, {Haynes}, {Haynes}, {Heiter}, {Helmi}, {Aguayo}, {Hewett}, {Hinton}, {Hobbs}, {Hoenig}, {Hofman}, {Hook}, {Hopgood}, {Hopkins}, {Hourihane}, {Howes}, {Howlett}, {Huet}, {Irwin}, {Iwert}, {Jablonka}, {Jahn}, {Jahnke}, {Jarno}, {Jin}, {Jofre}, {Johl}, {Jones}, {J{\"o}nsson}, {Jordan}, {Karovicova}, {Khalatyan}, {Kelz}, {Kennicutt}, {King}, {Kitaura}, {Klar}, {Klauser}, {Kneib}, {Koch}, {Koposov}, {Kordopatis}, {Korn}, {Kosmalski}, {Kotak}, {Kovalev}, {Kreckel}, {Kripak}, {Krumpe},
  {Kuijken}, {Kunder}, {Kushniruk}, {Lam}, {Lamer}, {Laurent}, {Lawrence}, {Lehmitz}, {Lemasle}, {Lewis}, {Li}, {Lidman}, {Lind}, {Liske}, {Lizon}, {Loveday}, {Ludwig}, {McDermid}, {Maguire}, {Mainieri}, {Mali}, \& {Mandel}}]{deJong2019}
{de Jong}, R.~S., {Agertz}, O., {Berbel}, A.~A., {et~al.} 2019, The Messenger, 175, 3

\bibitem[{{Eschen} {et~al.}(2024){Eschen}, {Bayliss}, {Wilson}, {Kunimoto}, {Pelisoli}, \& {Rodel}}]{Eschen2024}
{Eschen}, Y. N.~E., {Bayliss}, D., {Wilson}, T.~G., {et~al.} 2024, \mnras, 535, 1778

\bibitem[{{Gilbert} {et~al.}(2020){Gilbert}, {Barclay}, {Schlieder}, {Quintana}, {Hord}, {Kostov}, {Lopez}, {Rowe}, {Hoffman}, {Walkowicz}, {Silverstein}, {Rodriguez}, {Vanderburg}, {Suissa}, {Airapetian}, {Clement}, {Raymond}, {Mann}, {Kruse}, {Lissauer}, {Col{\'o}n}, {Kopparapu}, {Kreidberg}, {Zieba}, {Collins}, {Quinn}, {Howell}, {Ziegler}, {Vrijmoet}, {Adams}, {Arney}, {Boyd}, {Brande}, {Burke}, {Cacciapuoti}, {Chance}, {Christiansen}, {Covone}, {Daylan}, {Dineen}, {Dressing}, {Essack}, {Fauchez}, {Galgano}, {Howe}, {Kaltenegger}, {Kane}, {Lam}, {Lee}, {Lewis}, {Logsdon}, {Mandell}, {Monsue}, {Mullally}, {Mullally}, {Paudel}, {Pidhorodetska}, {Plavchan}, {Reyes}, {Rinehart}, {Rojas-Ayala}, {Smith}, {Stassun}, {Tenenbaum}, {Vega}, {Villanueva}, {Wolf}, {Youngblood}, {Ricker}, {Vanderspek}, {Latham}, {Seager}, {Winn}, {Jenkins}, {Bakos}, {Brice{\~n}o}, {Ciardi}, {Cloutier}, {Conti}, {Couperus}, {Di Sora}, {Eisner}, {Everett}, {Gan}, {Hartman}, {Henry}, {Isopi}, {Jao}, {Jensen}, {Law}, {Mallia}, {Matson},
  {Shappee}, {Le Wood}, \& {Winters}}]{Gilbert2020}
{Gilbert}, E.~A., {Barclay}, T., {Schlieder}, J.~E., {et~al.} 2020, \aj, 160, 116

\bibitem[{{Gilbert} {et~al.}(2023){Gilbert}, {Vanderburg}, {Rodriguez}, {Hord}, {Clement}, {Barclay}, {Quintana}, {Schlieder}, {Kane}, {Jenkins}, {Twicken}, {Kunimoto}, {Vanderspek}, {Arney}, {Charbonneau}, {G{\"u}nther}, {Huang}, {Isopi}, {Kostov}, {Kristiansen}, {Latham}, {Mallia}, {Mamajek}, {Mireles}, {Quinn}, {Ricker}, {Schulte}, {Seager}, {Suissa}, {Winn}, {Youngblood}, \& {Zapparata}}]{Gilbert2023}
{Gilbert}, E.~A., {Vanderburg}, A., {Rodriguez}, J.~E., {et~al.} 2023, \apjl, 944, L35

\bibitem[{{Goupil} {et~al.}(2024){Goupil}, {Catala}, {Samadi}, {Belkacem}, {Ouazzani}, {Reese}, {Appourchaux}, {Mathur}, {Cabrera}, {B{\"o}rner}, {Paproth}, {Moedas}, {Verma}, {Lebreton}, {Deal}, {Ballot}, {Chaplin}, {Christensen-Dalsgaard}, {Cunha}, {Lanza}, {Miglio}, {Morel}, {Serenelli}, {Mosser}, {Creevey}, {Moya}, {Garcia}, {Nielsen}, \& {Hatt}}]{Goupil2024}
{Goupil}, M.~J., {Catala}, C., {Samadi}, R., {et~al.} 2024, \aap, 683, A78

\bibitem[{{Heller} {et~al.}(2022){Heller}, {Harre}, \& {Samadi}}]{Heller2022}
{Heller}, R., {Harre}, J.-V., \& {Samadi}, R. 2022, \aap, 665, A11

\bibitem[{{Howell} {et~al.}(2014){Howell}, {Sobeck}, {Haas}, {Still}, {Barclay}, {Mullally}, {Troeltzsch}, {Aigrain}, {Bryson}, {Caldwell}, {Chaplin}, {Cochran}, {Huber}, {Marcy}, {Miglio}, {Najita}, {Smith}, {Twicken}, \& {Fortney}}]{Howell2014}
{Howell}, S.~B., {Sobeck}, C., {Haas}, M., {et~al.} 2014, \pasp, 126, 398

\bibitem[{{Kasting} {et~al.}(1993){Kasting}, {Whitmire}, \& {Reynolds}}]{Kasting1993}
{Kasting}, J.~F., {Whitmire}, D.~P., \& {Reynolds}, R.~T. 1993, \icarus, 101, 108

\bibitem[{{Kopparapu} {et~al.}(2013){Kopparapu}, {Ramirez}, {Kasting}, {Eymet}, {Robinson}, {Mahadevan}, {Terrien}, {Domagal-Goldman}, {Meadows}, \& {Deshpande}}]{Kopparapu2013}
{Kopparapu}, R.~K., {Ramirez}, R., {Kasting}, J.~F., {et~al.} 2013, \apj, 765, 131

\bibitem[{{Lebreton} \& {Goupil}(2014)}]{Lebreton2014}
{Lebreton}, Y. \& {Goupil}, M.~J. 2014, \aap, 569, A21

\bibitem[{{Lund} {et~al.}(2025){Lund}, {Chontos}, {Grundahl}, {Mathur}, {Garc{\'\i}a}, {Huber}, {Buzasi}, {Bedding}, {Hon}, \& {Li}}]{Lund2025}
{Lund}, M.~N., {Chontos}, A., {Grundahl}, F., {et~al.} 2025, \aap, 701, A285

\bibitem[{{Matuszewski} {et~al.}(2023){Matuszewski}, {Nettelmann}, {Cabrera}, {B{\"o}rner}, \& {Rauer}}]{Matuszewski2023}
{Matuszewski}, F., {Nettelmann}, N., {Cabrera}, J., {B{\"o}rner}, A., \& {Rauer}, H. 2023, \aap, 677, A133

\bibitem[{{Mayor} \& {Queloz}(1995)}]{Mayor1995}
{Mayor}, M. \& {Queloz}, D. 1995, \nat, 378, 355

\bibitem[{{Montalto} {et~al.}(2021){Montalto}, {Piotto}, {Marrese}, {Nascimbeni, V.}, {Prisinzano, L.}, {Granata, V.}, {Marinoni, S.}, {Desidera, S.}, {Ortolani, S.}, {Aerts, C.}, {Alei, E.}, {Altavilla, G.}, {Benatti, S.}, {B\"orner, A.}, {Cabrera, J.}, {Claudi, R.}, {Deleuil, M.}, {Fabrizio, M.}, {Gizon, L.}, {Goupil, M. J.}, {Heras, A. M.}, {Magrin, D.}, {Malavolta, L.}, {Mas-Hesse, J. M.}, {Pagano, I.}, {Paproth, C.}, {Pertenais, M.}, {Pollacco, D.}, {Ragazzoni, R.}, {Ramsay, G.}, {Rauer, H.}, \& {Udry, S.}}]{Montalto2021}
{Montalto}, M., {Piotto}, G., {Marrese}, P.~M., {et~al.} 2021, A\&A, 653, A98

\bibitem[{{Montalto} {et~al.}(2026){Montalto}, {Piotto}, {Marrese}, {Prisinzano}, {Marinoni}, {Granata}, {Cabrera}, {Nascimbeni}, {Desidera}, {Adibekyan}, {Ortolani}, {Alei}, {Aerts}, {Altavilla}, {Belkacem}, {Benatti}, {B{\"o}rner}, {Deleuil}, {Fabrizio}, {Gizon}, {Goupil}, {G{\"u}nther}, {Heras}, {Magrin}, {Malavolta}, {Mas-Hesse}, {Pagano}, {Paproth}, {Pollacco}, {Ragazzoni}, {Ramsay}, {Rauer}, \& {Udry}}]{Montalto2026}
{Montalto}, M., {Piotto}, G., {Marrese}, P.~M., {et~al.} 2026, arXiv e-prints, arXiv:2604.03369

\bibitem[{{Mowlavi} {et~al.}(2024){Mowlavi}, {Udry}, {Alonso}, {Bouchy}, {Desidera}, {Pollacco}, \& {Reiners}}]{Mowlavi2024}
{Mowlavi}, N., {Udry}, S., {Alonso}, R., {et~al.} 2024, in EAS2024, European Astronomical Society Annual Meeting, 1896

\bibitem[{{Nascimbeni} {et~al.}(2022){Nascimbeni}, {Piotto}, {B{\"o}rner}, {Montalto}, {Marrese}, {Cabrera}, {Marinoni}, {Aerts}, {Altavilla}, {Benatti}, {Claudi}, {Deleuil}, {Desidera}, {Fabrizio}, {Gizon}, {Goupil}, {Granata}, {Heras}, {Magrin}, {Malavolta}, {Mas-Hesse}, {Ortolani}, {Pagano}, {Pollacco}, {Prisinzano}, {Ragazzoni}, {Ramsay}, {Rauer}, \& {Udry}}]{Nascimbeni2022}
{Nascimbeni}, V., {Piotto}, G., {B{\"o}rner}, A., {et~al.} 2022, \aap, 658, A31

\bibitem[{{Nascimbeni} {et~al.}(2025){Nascimbeni}, {Piotto}, {Cabrera}, {Montalto}, {Marinoni}, {Marrese}, {Aerts}, {Altavilla}, {Benatti}, {B{\"o}rner}, {Deleuil}, {Desidera}, {Gizon}, {Goupil}, {Granata}, {Heras}, {Magrin}, {Malavolta}, {Mas-Hesse}, {Osborn}, {Pagano}, {Paproth}, {Pollacco}, {Prisinzano}, {Ragazzoni}, {Ramsay}, {Rauer}, {Tkachenko}, \& {Udry}}]{Nascimbeni2025}
{Nascimbeni}, V., {Piotto}, G., {Cabrera}, J., {et~al.} 2025, \aap, 694, A313

\bibitem[{{Pecaut} \& {Mamajek}(2013)}]{Pecaut2013}
{Pecaut}, M.~J. \& {Mamajek}, E.~E. 2013, \apjs, 208, 9

\bibitem[{{Prisinzano} {et~al.}(2026){Prisinzano}, {Montalto}, {Piotto}, {Marrese}, {Marinoni}, {Nascimbeni}, {Granata}, {Cabrera}, {Belkacem}, {Deleuil}, {Gizon}, {Goupil}, {Pagano}, {Pollacco}, {Ragazzoni}, {Rauer}, {Udry}, {Maldonado}, {Micela}, {Damiani}, {Affer}, {Altavilla}, {Argiroffi}, {Benatti}, {Cassisi}, {Claudi}, {Desidera}, {Fabrizio}, {Flaccomio}, {Heiter}, {Lanza}, {Maggio}, {Malavolta}, {Nardiello}, {Ortolani}, \& {Sozzetti}}]{Prisinzano2026}
{Prisinzano}, L., {Montalto}, M., {Piotto}, G., {et~al.} 2026, \aap, 706, A207

\bibitem[{{Rauer} {et~al.}(2025){Rauer}, {Aerts}, {Cabrera}, {Deleuil}, {Erikson}, {Gizon}, {Goupil}, {Heras}, {Walloschek}, {Lorenzo-Alvarez}, {Marliani}, {Martin-Garcia}, {Mas-Hesse}, {O'Rourke}, {Osborn}, {Pagano}, {Piotto}, {Pollacco}, {Ragazzoni}, {Ramsay}, {Udry}, {Appourchaux}, {Benz}, {Brandeker}, {G{\"u}del}, {Janot-Pacheco}, {Kabath}, {Kjeldsen}, {Min}, {Santos}, {Smith}, {Suarez}, {Werner}, {Aboudan}, {Abreu}, {Acu{\~n}a}, {Adams}, {Adibekyan}, {Affer}, {Agneray}, {Agnor}, {Aguirre B{\o}rsen-Koch}, {Ahmed}, {Aigrain}, {Al-Bahlawan}, {Alcacera Gil}, {Alei}, {Alencar}, {Alexander}, {Alfonso-Garz{\'o}n}, {Alibert}, {Allende Prieto}, {Almeida}, {Alonso Sobrino}, {Altavilla}, {Althaus}, {Alvarez Trujillo}, {Amarsi}, {Ammler-von Eiff}, {Am{\^o}res}, {Andrade}, {Antoniadis-Karnavas}, {Ant{\'o}nio}, {Aparicio del Moral}, {Appolloni}, {Arena}, {Armstrong}, {Aroca Aliaga}, {Asplund}, {Audenaert}, {Auricchio}, {Avelino}, {Baeke}, {Bailli{\'e}}, {Balado}, {Ballber Balaguer{\'o}}, {Balestra}, {Ball}, {Ballans},
  {Ballot}, {Barban}, {Barbary}, {Barbieri}, {Barcel{\'o} Forteza}, {Barker}, {Barklem}, {Barnes}, {Barrado Navascues}, {Barragan}, {Baruteau}, {Basu}, {Baudin}, {Baumeister}, {Bayliss}, {Bazot}, {Beck}, {Belkacem}, {Bellinger}, {Benatti}, {Benomar}, {B{\'e}rard}, {Bergemann}, {Bergomi}, {Bernardo}, {Biazzo}, {Bignamini}, {Bigot}, {Billot}, {Binet}, {Biondi}, {Biondi}, {Birch}, {Bitsch}, {Bluhm Ceballos}, {B{\'o}di}, {Bogn{\'a}r}, {Boisse}, {Bolmont}, {Bonanno}, {Bonavita}, {Bonfanti}, {Bonfils}, {Bonito}, {Bonomo}, {B{\"o}rner}, {Boro Saikia}, {Borreguero Mart{\'\i}n}, {Borsa}, {Borsato}, {Bossini}, {Bouchy}, {Bou{\'e}}, {Boufleur}, {Boumier}, {Bourrier}, {Bowman}, {Bozzo}, {Bradley}, {Bray}, {Bressan}, {Breton}, {Brienza}, {Brito}, {Brogi}, {Brown}, {Brown}, {Brun}, {Bruno}, {Bruns}, {Buchhave}, {Bugnet}, {Buldgen}, {Burgess}, {Busatta}, {Busso}, {Buzasi}, {Caballero}, {Cabral}, {Cabrero Gomez}, {Calderone}, {Cameron}, {Cameron}, {Campante}, {Campos Gestal}, {Canto Martins}, {Cara}, {Carone}, {Carrasco},
  {Casagrande}, {Casewell}, {Cassisi}, {Castellani}, {Castro}, {Catala}, {Catal{\'a}n Fern{\'a}ndez}, {Catelan}, {Cegla}, {Cerruti}, {Cessa}, {Chadid}, {Chaplin}, {Charpinet}, {Chiappini}, {Chiarucci}, {Chiavassa}, {Chinellato}, {Chirulli}, {Christensen-Dalsgaard}, {Church}, {Claret}, {Clarke}, {Claudi}, {Clermont}, {Coelho}, {Coelho}, {Cogato}, {Colom{\'e}}, {Condamin}, {Conde Garc{\'\i}a}, \& {Conseil}}]{Rauer2025}
{Rauer}, H., {Aerts}, C., {Cabrera}, J., {et~al.} 2025, Experimental Astronomy, 59, 26

\bibitem[{{Ricker} {et~al.}(2015){Ricker}, {Winn}, {Vanderspek}, {Latham}, {Bakos}, {Bean}, {Berta-Thompson}, {Brown}, {Buchhave}, {Butler}, {Butler}, {Chaplin}, {Charbonneau}, {Christensen-Dalsgaard}, {Clampin}, {Deming}, {Doty}, {De Lee}, {Dressing}, {Dunham}, {Endl}, {Fressin}, {Ge}, {Henning}, {Holman}, {Howard}, {Ida}, {Jenkins}, {Jernigan}, {Johnson}, {Kaltenegger}, {Kawai}, {Kjeldsen}, {Laughlin}, {Levine}, {Lin}, {Lissauer}, {MacQueen}, {Marcy}, {McCullough}, {Morton}, {Narita}, {Paegert}, {Palle}, {Pepe}, {Pepper}, {Quirrenbach}, {Rinehart}, {Sasselov}, {Sato}, {Seager}, {Sozzetti}, {Stassun}, {Sullivan}, {Szentgyorgyi}, {Torres}, {Udry}, \& {Villasenor}}]{Ricker2015}
{Ricker}, G.~R., {Winn}, J.~N., {Vanderspek}, R., {et~al.} 2015, Journal of Astronomical Telescopes, Instruments, and Systems, 1, 014003

\bibitem[{{Smith} {et~al.}(2025){Smith}, {Ahmed}, {De Angeli}, {Burgess}, {Busso}, {Ford}, {Harrison}, {Hodgkin}, {Irwin}, {Rixon}, \& {Walton}}]{Smith2025}
{Smith}, L.~C., {Ahmed}, S., {De Angeli}, F., {et~al.} 2025, \mnras, 539, 297

\bibitem[{{Soderblom} \& {King}(1998)}]{Soderblom1998}
{Soderblom}, D.~R. \& {King}, J.~R. 1998, in Solar Analogs: Characteristics and Optimum Candidates., ed. J.~C. {Hall}, 41

\bibitem[{{Stassun} {et~al.}(2018){Stassun}, {Oelkers}, {Pepper}, {Paegert}, {De Lee}, {Torres}, {Latham}, {Charpinet}, {Dressing}, {Huber}, {Kane}, {L{\'e}pine}, {Mann}, {Muirhead}, {Rojas-Ayala}, {Silvotti}, {Fleming}, {Levine}, \& {Plavchan}}]{Stassun2018}
{Stassun}, K.~G., {Oelkers}, R.~J., {Pepper}, J., {et~al.} 2018, \aj, 156, 102

\bibitem[{{Tayar} {et~al.}(2022){Tayar}, {Claytor}, {Huber}, \& {van Saders}}]{Tayar2022}
{Tayar}, J., {Claytor}, Z.~R., {Huber}, D., \& {van Saders}, J. 2022, \apj, 927, 31

\bibitem[{{Teske} {et~al.}(2021){Teske}, {Wang}, {Wolfgang}, {Gan}, {Plotnykov}, {Armstrong}, {Butler}, {Cale}, {Crane}, {Howard}, {Jensen}, {Law}, {Shectman}, {Plavchan}, {Valencia}, {Vanderburg}, {Ricker}, {Vanderspek}, {Latham}, {Seager}, {Winn}, {Jenkins}, {Adibekyan}, {Barrado}, {Barros}, {Benkhaldoun}, {Brown}, {Bryant}, {Burt}, {Caldwell}, {Charbonneau}, {Cloutier}, {Collins}, {Collins}, {Colon}, {Conti}, {Demangeon}, {Eastman}, {Elmufti}, {Feng}, {Flowers}, {Guerrero}, {Hojjatpanah}, {Irwin}, {Isopi}, {Lillo-Box}, {Mallia}, {Massey}, {Mori}, {Mullally}, {Narita}, {Nishiumi}, {Osborn}, {Paegert}, {de Leon}, {Quinn}, {Reefe}, {Schwarz}, {Shporer}, {Soubkiou}, {Sousa}, {Stockdale}, {Str{\o}m}, {Tan}, {Tang}, {Tenenbaum}, {Wheatley}, {Wittrock}, {Yahalomi}, \& {Zohrabi}}]{Teske2021}
{Teske}, J., {Wang}, S.~X., {Wolfgang}, A., {et~al.} 2021, \apjs, 256, 33

\bibitem[{{Trifonov} {et~al.}(2019){Trifonov}, {Rybizki}, \& {K{\"u}rster}}]{Trifonov2019}
{Trifonov}, T., {Rybizki}, J., \& {K{\"u}rster}, M. 2019, \aap, 622, L7

\bibitem[{{Vaulato} {et~al.}(2025){Vaulato}, {Hobson}, {Allart}, {Pelletier}, {Wardenier}, {Chakraborty}, {Ehrenreich}, {Nari}, {Steiner}, {Dumusque}, {Hoeijmakers}, {Artigau}, {Baron}, {Barros}, {Benneke}, {Bonfils}, {Bouchy}, {Bryan}, {Canto Martins}, {Cloutier}, {Cook}, {Cowan}, {De Medeiros}, {Delfosse}, {Delgado-Mena}, {Doyon}, {Gonz{\'a}lez Hern{\'a}ndez}, {Lafreni{\`e}re}, {de Castro Le{\~a}o}, {Lovis}, {Malo}, {Melo}, {Mignon}, {Mordasini}, {Pepe}, {Rebolo}, {Rowe}, {Santos}, {S{\'e}gransan}, {Su{\'a}rez Mascare{\~n}o}, {Udry}, {Valencia}, {Wade}, {Aguiar}, {Al Moulla}, {Akinsanmi}, {Borsato}, {Cadieux}, {Carteret}, {Costa Silva}, {Cristo}, {Forveille}, {Frensch}, {Gromek}, {Lendl}, {Prinoth}, {Psaridi}, {Stefanov}, {Thorsbro}, \& {Weisserman}}]{Vaulato2025}
{Vaulato}, V., {Hobson}, M.~J., {Allart}, R., {et~al.} 2025, \aap, 703, A251

\bibitem[{{Walcher} {et~al.}(2019){Walcher}, {Banerji}, {Battistini}, {Bell}, {Bellido-Tirado}, {Bensby}, {Bestenlehner}, {Boller}, {Brynnel}, {Casey}, {Chiappini}, {Christlieb}, {Church}, {Cioni}, {Croom}, {Comparat}, {Davies}, {de Jong}, {Dwelly}, {Enke}, {Feltzing}, {Feuillet}, {Fouesneau}, {Ford}, {Frey}, {Gonzalez-Solares}, {Gueguen}, {Howes}, {Irwin}, {Klar}, {Kordopatis}, {Korn}, {Krumpe}, {Kushniruk}, {Lam}, {Lewis}, {Lind}, {Liske}, {Loveday}, {Mainieri}, {Martell}, {Matijevic}, {McMahon}, {Merloni}, {Murphy}, {Niederhofer}, {Norberg}, {Pramskiy}, {Romaniello}, {Robotham}, {Rothmaier}, {Ruchti}, {Schnurr}, {Schwope}, {Smedley}, {Sorce}, {Starkenburg}, {Stilz}, {Storm}, {Tempel}, {Thi}, {Traven}, {Valentini}, {van den Ancker}, {Walton}, {Winkler}, \& {Worley}}]{Walcher2019}
{Walcher}, C.~J., {Banerji}, M., {Battistini}, C., {et~al.} 2019, The Messenger, 175, 12

\bibitem[{{Winn}(2010)}]{Winn2010}
{Winn}, J.~N. 2010, in Exoplanets, ed. S.~{Seager}, 55--77

\bibitem[{{Wolszczan} \& {Frail}(1992)}]{Wolszczan1992}
{Wolszczan}, A. \& {Frail}, D.~A. 1992, \nat, 355, 145

\end{thebibliography}


\end{document}